\documentclass[a4paper,onecolumn,11pt,accepted=2022-10-22]{quantumarticle}
\pdfoutput=1
\usepackage[utf8]{inputenc}
\usepackage[english]{babel}
\usepackage[T1]{fontenc}
\usepackage{amsmath}
\usepackage{hyperref}
\usepackage{amsthm}
\usepackage{latexsym}
\usepackage{amssymb}
\usepackage{bbm,dsfont}
\usepackage{enumerate}

\newtheorem{theorem}{Theorem}[section]
\newtheorem{proposition}[theorem]{Proposition}

\newtheorem{corollary}[theorem]{Corollary}

\theoremstyle{definition}

\newtheorem{remark}[theorem]{Remark}
\newtheorem{definition}[theorem]{Definition}
\newtheorem{example}[theorem]{Example}

\newcommand{\real}{\mathbb R} %real
\newcommand{\integer}{\mathbb Z} %integer
\newcommand{\hi}{\mathcal{H}} %Hilbert space H
\newcommand{\lh}{\mathcal{L(H)}} %bounded linear operators
\newcommand{\lhs}{\mathcal{L}_s(\hi)} %selfadjoint bounded linear operators
\newcommand{\lhp}{\mathcal{L}_+(\hi)} %positive operators
\newcommand{\sh}{\mathcal{S(H)}} %states on H
\newcommand{\eh}{\mathcal{E(H)}} %effects
\newcommand{\ip}[2]{\left\langle\,#1\,|\,#2\,\right\rangle} %inner product
\newcommand{\ket}[1]{|#1\rangle} %ket
\newcommand{\bra}[1]{\langle#1|} %bra
\newcommand{\ketbra}[2]{|#1\rangle\langle#2|} %ketbra
\newcommand{\tr}[1]{\mathrm{tr}\left[#1\right]} %trace
\newcommand{\supp}[1]{\mathrm{supp}\left(#1\right)} %support
\newcommand{\id}{\mathbbm{1}} %identity operator
\newcommand{\Fou}{\mathcal{F}} %Fourier

\newcommand{\obs}{\mathcal{O}}
\newcommand{\obsh}{\mathcal{O}_\hi}
\newcommand{\obsd}{\mathcal{O}_d}
\newcommand{\obsdsim}{\mathcal{O}_d^\sim}
\newcommand{\A}{\mathsf{A}}%generic observable
\newcommand{\B}{\mathsf{B}}%generic observable
\newcommand{\C}{\mathsf{C}}%generic observable
\newcommand{\M}{\mathsf{M}}%joint obs
\newcommand{\Y}{\mathsf{Y}}%generic observable

\newcommand{\T}{\mathsf{T}}

\newcommand{\pleq}{\preccurlyeq} %pp order
\newcommand{\pgeq}{\succcurlyeq} %pp order

\newcommand{\fim}{f}
\newcommand{\fimrho}{f_\rho}
\newcommand{\Fim}{F}
\newcommand{\Fimrho}{F_\rho}

\author[Heinosaari]{Teiko Heinosaari}
\affiliation{Quantum algorithms and software, VTT Technical Research Centre of Finland Ltd}
\affiliation{Department of Physics and Astronomy, University of Turku, Finland}
\email{teiko.heinosaari@utu.fi}
\orcid{0000-0003-2405-5439}

\author[Jivulescu]{Maria Anastasia Jivulescu}
\affiliation{Department of Mathematics,
Politehnica University of Timi\c soara, Romania}
\email{maria.jivulescu@upt.ro}
\orcid{0000-0001-6200-5630}

\author[Nechita]{Ion Nechita}
\affiliation{Laboratoire de Physique Th\'eorique, Universit\'e de Toulouse, CNRS, UPS, France}
\email{ion.nechita@univ-tlse3.fr}
\orcid{0000-0003-3016-7795}

\begin{document}

\title[Order preserving maps on quantum measurements]{Order preserving maps on quantum measurements}

\begin{abstract} 
We study the partially ordered set of equivalence classes of quantum measurements endowed with the post-processing partial order. 
The post-processing order is fundamental as it enables to compare measurements by their intrinsic noise and it gives grounds to define the important concept of quantum incompatibility. 
Our approach is based on mapping this set into a simpler partially ordered set using an order preserving map and investigating the resulting image. 
The aim is to ignore unnecessary details while keeping the essential structure, thereby simplifying e.g. detection of incompatibility.
One possible choice is the map based on Fisher information introduced by Huangjun Zhu, known to be an order morphism taking values in the cone of positive semidefinite matrices. 
We explore the properties of that construction and improve Zhu's incompatibility criterion 
by adding a constraint depending on the number of measurement outcomes. 
We generalize this type of construction to other ordered vector spaces and we show that 
this map is optimal among all quadratic maps.
\end{abstract}

\maketitle

\tableofcontents

%%%%%%%%%%%%%
\section{Introduction}
%%%%%%%%%%%%%

There is a general guideline in mathematics how to study a complicated structure: map it to something simpler while keeping the essential structural features of it. Then, by investigating the simple structure one can get information also on the more complicated structure.
In the present work, we follow this guideline to investigate the post-processing order of quantum measurements.
Post-processing can be defined as a classical manipulation of measurement outcome data, but it can also be arising e.g.~from weak measurement coupling, disturbance caused by environment or some other source of noise. Therefore, post-processing ordering does not mean that there has been active processing of measurement data, but rather that in principle two measurements could be related in that specific way.
Generally, one can say that the post-processing order describes which measurement is noisier than another measurement.
An important aspect is that the post-processing order provides an operational definition of quantum incompatibility, which has become central in several quantum information processing applications. The post-processing relation was initially studied in \cite{MaMu90a} and further results and developments have been reported e.g. in \cite{Heinonen05, BuDaKePeWe05, JePu07, JePuVi08, AlCaHeTo09,HaHePe12,Kuramochi15a, Kuramochi15b, HaPe17, GuMcSaGi21}.

The starting point of our investigation is that we map the set of all quantum measurements to a partially ordered vector space by an order preserving map. We demonstrate how quadratic maps on matrices lead to this kind of setting in a natural way.
This approach has been used by Huangjun Zhu \cite{Zhu15,ZhHaCh16}, who defined a particular kind of map related to Fisher information. We show that the map introduced by Zhu is exceptional among all quadratic maps and we explore the properties of that specific construction.
We also give examples of other order preserving maps and show that some of them can be useful when investigating the post-processing order of quantum measurements, although they cannot detect incompatibility. 
The simpler the map is, the less information it gives. But a complicated map does not simplify the structure enough and hence does not provide any actual relief. 
From this respect, Zhu's map is in good balance and that explains its special status and usefulness for concrete calculations.

The paper is organized as follows.
In Section \ref{sec:preorder} we explain the core idea in a general setting of preordered sets. 
In the rest of the paper we then limit to quantum measurements. Their post-processing order is recalled Section \ref{sec:compatibility}.
Then, in Section \ref{sec:order-maps} we explain a method to construct order preserving maps on the set of quantum measurements and we provide several examples. 
After presenting the general framework, in Section~\ref{sec:FI-map} we go into the details of Zhu's construction. In Section~\ref{sec:incompatibility-criterion} we recall and improve Zhu's incompatibility criterion by adding a constraint depending on the number of outcomes of the quantum measurement. Finally, in Section~\ref{sec:examples} we provide several examples that illustrate the improved Zhu criterion for incompatibility.

%%%%%%%%%%%%%%%%%%%%%%
\section{Prelude: (in)compatibility in preordered sets}\label{sec:preorder}
%%%%%%%%%%%%%%%%%%%%%%

To get a structural perspective to post-processing and incompatibility, we start with a general setting of preordered sets.

\begin{figure}
\centering
 \includegraphics[scale=0.15]{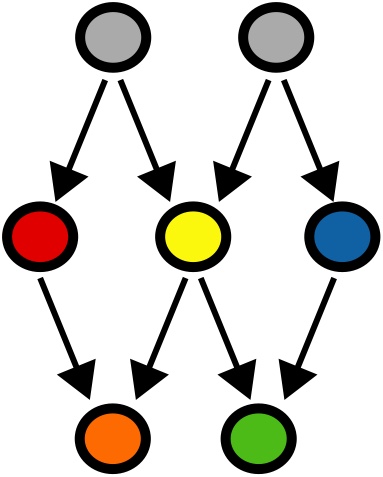}\qquad\qquad\qquad 
    \includegraphics[scale=0.15]{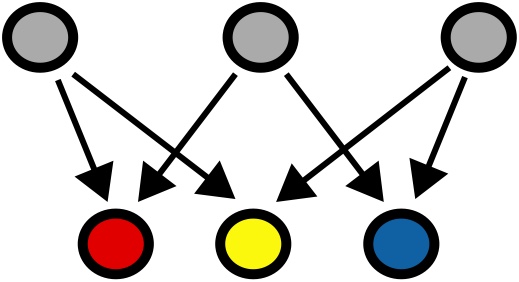}
    \qquad\qquad\qquad 
    \includegraphics[scale=0.15]{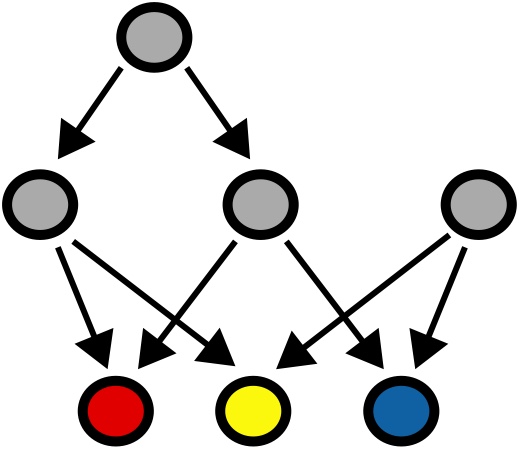}
\caption{From left to right: (a) All six colored elements are pairwise compatible except the pair consisting red and blue elements. (b) All colored elements are pairwise compatible but the colored triplet is incompatible. (c) The colored triplet is compatible.}
\label{fig:graph}
\end{figure}

We recall that a \emph{preorder} on a set $\Omega$ is a binary relation $\leq$ that is reflexive and transitive.
For a nonempty subset $X\subseteq \Omega$, we denote
\begin{align*}
\uparrow X &= \{ y \in \Omega : x \leq y \quad \forall x \in X\} \, , \\
\downarrow X &= \{ y \in \Omega : y \leq x  \quad \forall x \in X\} \, .
\end{align*}
A subset $X$ is called \emph{compatible} if $\uparrow X \neq \emptyset$, i.e., there exists an element $y\in \Omega$ such that  $x \leq y$ for all $x\in X$; otherwise $X$ is called \emph{incompatible}.
We also say that the elements in $X$ are (in)compatible when $X$ is (in)compatible.
A preorderer set can be visualized as a directed graph, with elements of the set corresponding to vertices while the preorder between two elements corresponds to the directed edges between vertices.
Illustrations are given in Fig.~\ref{fig:graph}. 

The operational interpretation of compatibility and incompatibility depends on the preorder.  
In this section, we make some general definitions and we further illustrate the concept with some simple examples.
Our primary interest, namely, the post-processing preorder of quantum measurements, is defined later in Sec.~\ref{sec:compatibility}.  

We recall the following notions.
An element $x\in \Omega$ is 
\begin{itemize}
\item \emph{maximal} if $x\leq y$ implies that $y\leq x$ for all $y\in \Omega$.
\item \emph{minimal} if $y \leq x$ implies that $x\leq y$ for all $y\in \Omega$.
\end{itemize}
There can be many maximal and minimal elements (see Fig.~\ref{fig:graph}), although in exceptional cases they are unique and then called the greatest and the least element, respectively. We note that two maximal elements $x$ and $y$ are compatible if and only if $x\leq y \leq x$. 

\begin{example}\label{ex:convex}
Let $\Omega$ be a convex set. 
For two elements $x,y\in\Omega$, we define $x \leq y$ if $y$ is in some nontrivial convex decomposition of $x$, i.e., $x=t y + (1-t) y'$ for some $0<t<1$ and $y'\in\Omega$.
The maximal elements are, by definition, the extreme elements of $\Omega$. Any collection of different extreme elements is incompatible. 
\end{example}

We recall that a preorder $\leq$ is a \emph{partial order} if it is antisymmetric, i.e., $x \leq y \leq x$ implies $x=y$. 
The usual operator ordering, recalled in the next example, is a partial order and we will use it later.

\begin{example}
Let $\hi$ be a complex Hilbert space and $\lhp$ the set of all positive operators on $\hi$, where an operator $T$ is \emph{positive} if $\ip{\psi}{T\psi}\geq 0$ for all $\psi\in\hi$. Positive operators correspond to positive semidefinite matrices. The usual operator ordering is defined as $S\leq T$ if $T-S$ is positive, and it is a partial order. For any $S,T\in\lhp$, we have $S\leq S+T$ and $T \leq S+T$, hence $S+T$ is an upper bound for $S$ and $T$.
However, it is known that $\lhp$ is not a lattice but an antilattice; the least upper bound for $S$ and $T$ exists if and only if $S\leq T$ or $T\leq S$ \cite{Kadison51}.
\end{example}

For our purposes, there is no big difference in preorder and partial order. 
Namely, a preorder $\leq$ determines an equivalence relation $\simeq$ on $\Omega$ by setting $x\simeq y$ if  $x\leq y \leq x$.
We denote by $[x]$ the equivalence class of $x$.
Further, $\leq$ determines a partial order on the set of all equivalence classes as follows: $[x] \leq [y]$ if and only if $x \leq y$ (without causing confusion we denote the induced partial order with the same symbol $\leq$). We denote by $\Omega^\simeq$ this partially ordered set (poset).

Importantly, it follows from the definition of compatibility that $X\subseteq \Omega$ is compatible if and only if $X^{[]}\equiv \{[x]:x\in X\}$ is compatible.
We conclude that there is no essential difference if we investigate compatibility in a preordered set or in the induced partially ordered set.

\begin{example}\label{ex:prob-1}
Let $\mathcal{P}(n)$ be the convex set of probability distributions on the set $\{1,\ldots,n\}$.
We consider the preorder defined in Example \ref{ex:convex}.
We can interpret the Kronecker distributions $\delta_j$ as describing precise knowledge, whereas other probability distributions describe knowledge with some uncertainty. 
For $p,q\in\mathcal{P}(n)$, we have $p\leq q$ if and only if $\supp{p}\supseteq \supp{q}$.
We see that $p,q\in\mathcal{P}(n)$ are compatible if and only if there exists $j\in\{1,\ldots,n\}$ such that $p\leq \delta_j$ and $q\leq\delta_j$, which is the case when $p(j)q(j)\neq 0$.
Hence, compatibility of $p$ and $q$ means that they may origin from the same precise description $\delta_j$, just the uncertainty is different.
We have $p\simeq q$ if and only if $\supp{p}=\supp{q}$.
The maximal elements in $\mathcal{P}(n)^\simeq$ are the singleton sets $\{\delta_j\}$, $j=1,\ldots,n$, and the least element is the subset containing all probability distributions with full support.
\end{example}

The main tool in our later developments will be order preserving maps between two preordered (or partially ordered) sets.
Let $(\Omega,\leq)$ and $(\Omega',\leq')$ be two preordered sets.
A map $h:\Omega \to \Omega'$ is order preserving if
\begin{equation*}
x \leq y \quad \Rightarrow \quad h(x) \leq' h(y)
\end{equation*}
for all $x,y\in\Omega$.
We make the following simple observations.
\begin{proposition}\label{prop:f}
Let $h$ be an order preserving map. For any $X\subseteq\Omega$, the following implications hold:
\begin{enumerate}
\item[(a)] If $X$ is compatible, then the image $h(X)$ is compatible.
\item[(b)] If the image $h(X)$ is incompatible, then $X$ is incompatible.
\end{enumerate}
\end{proposition}

\begin{figure}
\centering
 \includegraphics[scale=0.15]{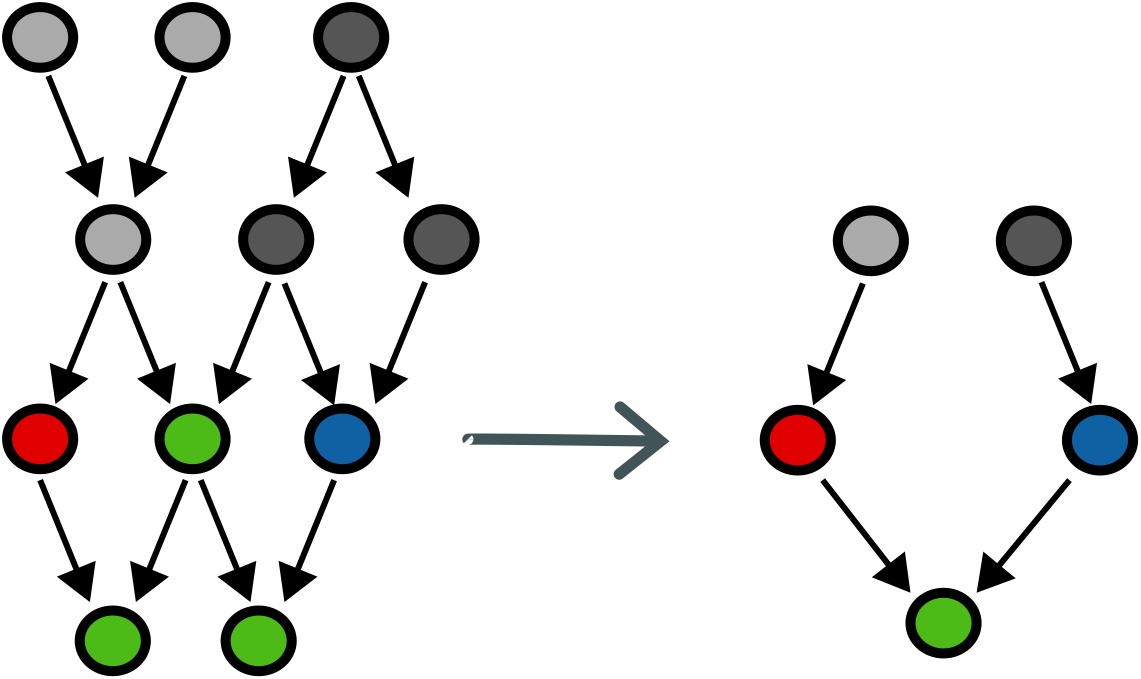}\qquad\qquad\qquad 
    \includegraphics[scale=0.15]{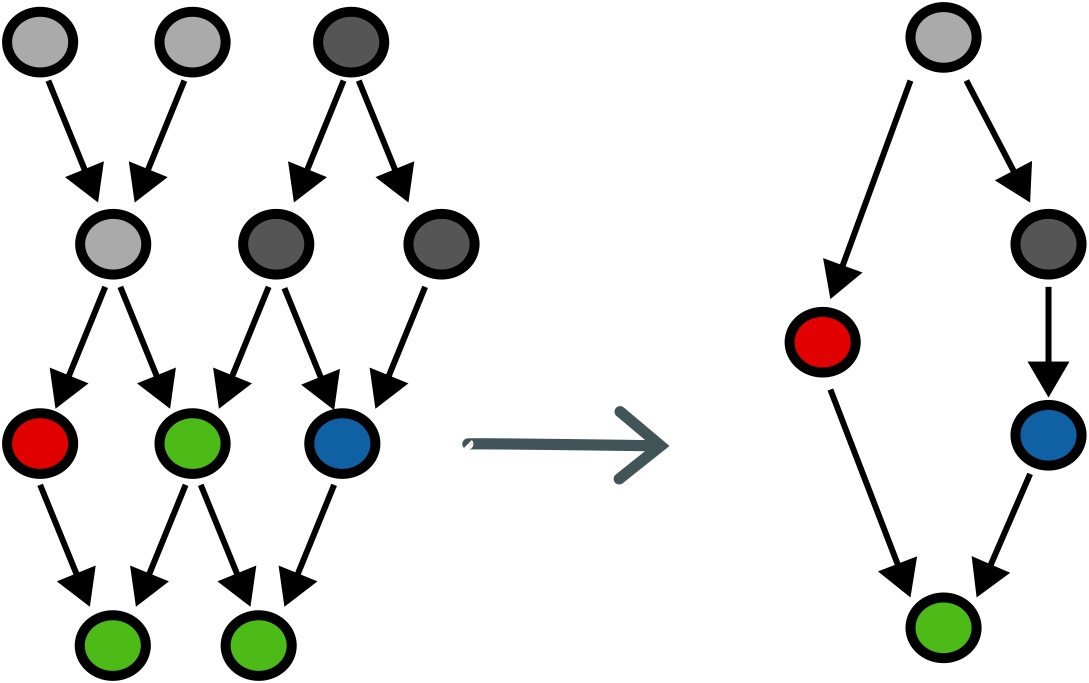}
\caption{A partially order set is mapped to another partially order set by an order preserving map. Similarly colored elements are mapped into respectively colored elements. From left to right: (a) In the image side the incompatibility of the red and blue elements is evident, hence they are incompatible also on the domain side. (b)  The image contains the greatest element. Therefore, all the elements on the image side are compatible and the map is useless for incompatibility detection. }
\label{fig:graph-incomp}
\end{figure}

The latter observation is the starting point for our investigation and it is illustrated in Fig.~\ref{fig:graph-incomp}(a).
This observation means that we can obtain sufficient criteria for incompatibility on $\Omega$ by investigating other preordered sets and relevant order preserving maps. 

Suppose we want to investigate a preordered set $(\Omega,\leq)$ and for that aim we choose  sets $(\Omega',\leq')$ and order preserving maps $h:\Omega \to \Omega'$. 
There is an obvious trade-off between lost information and difficulty. Namely, simpler is the range $h(\Omega)$, easier it is to decide if the image $h(X)$ of a subset $X$ is incompatible. In contrast, in a simpler case we loose more about the original structure of $(\Omega,\leq)$ and hence detect less incompatible sets. 

\begin{example}\label{ex:catastrophe}
Let $(\Omega,\leq)$ and $(\Omega',\leq')$ be two preordered sets and $h:\Omega \to \Omega'$ an order preserving map. 
Suppose that $h(\Omega)$ has the greatest element.
It follows that $h(X)$ is compatible for any $X\subset\Omega$ and Prop.~\ref{prop:f}(b) is therefore useless in detecting incompatibility. This kind of situation is depicted in Fig.~\ref{fig:graph-incomp}(b).
\end{example}

%%%%%%%%%%%%%
\section{Post-processing order of quantum measurements}\label{sec:compatibility}
%%%%%%%%%%%%%

We review in this section the basic theory of quantum measurements, using the formalism of Positive Operator Valued Measures (POVMs), also known as quantum meters. 
Our focus is on the partial ordering induced by \emph{classical post-processing}, and the related, crucial phenomenon of \emph{quantum incompatibility}. 

%%%%%%%%%%%%%
\subsection{Quantum measurements}
%%%%%%%%%%%%%

In the following $\hi$ is a fixed finite $d$-dimensional Hilbert space.
We denote by $\sh$ the set of all states, i.e., positive trace class operators of trace 1 and by  $\eh$ the set of effects, that are linear operators $E$ on $\hi$  such that $0\leq E \leq \id$.
We will restrict our investigation to quantum measurements with a finite number of outcomes, hence we will understand a measurement as a map $\A:x\to\A_x$ from a finite set of measurement outcomes $\Omega_\A$ to the set of bounded linear operators $\lh$ on $\hi$ such that $\A_x\geq 0$ and $\sum_x \A_x = \id$.
This is also known as a positive operator valued measure (POVM).
For a subset $X\subseteq\Omega_\A$, we denote $\A_X = \sum_{x\in X} \A_x$.
The probability of getting an outcome $x$ in a measurement of $\A$ in an initial state $\varrho$ is given by the formula $\tr{\varrho \A_x}$.
We denote by $\obsd$ the set of all measurements $\A$ on $\hi$ with a finite outcome set $\Omega_\A \subset \integer$.

%%%%%%%%%%%%%%%%%%%%%%%%%%%%%%%%%%%%%
\subsection{Post-processing preorder}
%%%%%%%%%%%%%%%%%%%%%%%%%%%%%%%%%%%

We recall in this section the post-processing order relation on the set of quantum measurements (or POVMs). A matrix $(\mu_{xy})_{x \in X, y \in Y}$ is called \emph{column stochastic} if it has non-negative elements and its column sums are 1: 
$$\forall y \in Y, \qquad \sum_{x \in X} \mu_{xy} = 1.$$

\begin{definition}\label{def:post-processing}
For two measurements $\A$ and $\B$, we denote $\A\pleq\B$ if there exists a column stochastic matrix $\mu$ such that
\begin{align}\label{eq:pp-basic}
\forall x \in X, \qquad \A_x = \sum_{y \in Y} \mu_{xy} \B_y \, .
\end{align}
We call this relation \emph{post-processing}.
\end{definition}

It is straightforward to verify that $\pleq$ is a preorder.

\begin{example}
A special type of post-processing is \emph{relabeling}, which means that $\mu_{xy}\in\{0,1\}$ for all $x,y$. In this case, $\mu$ being column stochastic implies that there is a relabeling function $r:X \to Y$ such that \eqref{eq:pp-basic} can be written as
\begin{align}\label{eq:relabeling}
\forall x \in X, \qquad \A_x=\sum_{y\in r^{-1}(x)} \B_y.
\end{align}
In this special case, we say that $\A$ is a relabeling of $\B$ and that $\B$ is a refinement of $\A$.
\end{example}

In the following, we summarize some known results that are relevant for our investigation. 
The proofs can be found in \cite{MaMu90a}.

\begin{proposition}
\begin{enumerate}[(a)]
\item A measurement $\A\in\obsd$ is post-processing minimal if and only if $\A$ is trivial, i.e.~ each $\A_x = a_x\id$ for some $a_x\in\real$.
\item A measurement $\A\in\obsd$ is post-processing maximal if and only if $\A$ is rank-1, i.e.~each nonzero $\A_x$ is a rank-1 operator.
\end{enumerate}
\end{proposition}

%%%%%%%%%%%%%%%%%%%
\subsection{Equivalence classes and simple representatives}\label{sec:simple}
%%%%%%%%%%%%%%%%%%%

The post-processing preorder relation introduced above is not anti-symmetric, meaning that there exist distinct measurements $\A \neq \B$ such that both $\A \pleq \B$ and $\B \pleq \A$ hold. To turn $\pleq$ into a partial order, we need to quotient it by the corresponding equivalence classes as explained in Section \ref{sec:preorder}. 

\begin{definition}
Given two measurements $\A$ and $\B$, we denote $\A\sim\B$ if $\A\pleq\B\pleq\A$, and in this case we say that $\A$ and $\B$ are \emph{post-processing equivalent}. We denote by $[\A]$ the equivalence class of an measurement $\A$, and by $\obsdsim$ the set of all equivalence classes of measurements on $\mathcal H \cong \mathbb C^d$.

The relation $\pleq$ induces a \emph{partial order} (which we also denote by $\pleq$) on the set $\obsdsim$ of equivalence classes of measurements. 
\end{definition}

\begin{figure}
\centering
    \includegraphics[scale=0.75]{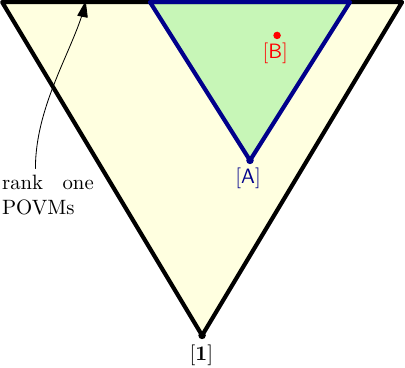}\qquad\qquad\qquad 
    \includegraphics[scale=0.75]{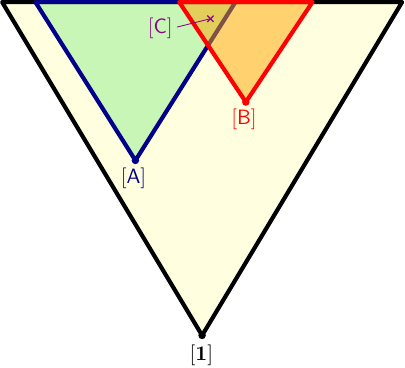}
    \caption{The poset of equivalence classes of quantum measurements has the least element $[\id]$ and infinitely many maximal elements. Left: $[\A] \prec [\B]$. Right: compatible measurements $\A$ and $\B$ with a joint measurement $\C$.} 
    \label{fig:obsdsim}
\end{figure} 

It is convenient to be able to choose a ``standard representative'' of an equivalence class $[\A]$ of measurements when needed. We say that a measurement $\A$ is \emph{simple} if every $\B\in[\A]$ is a refinement of $\A$. 
It has been shown in \cite{Kuramochi15b} that every equivalence class has a simple representative (called minimally sufficient in  \cite{Kuramochi15b}) that can be formed iteratively as follows: start from $\A$ and merge two operators $\A_x$ and $\A_y$ if there exists $c\geq0$ such that $\A_x=c\A_y$. Then continue this merging until there are no more such pairs.
It follows that a measurement $\A$ is simple if and only if, whenever $x,y\in\Omega_A$ and $x\neq y$, there is no $c \geq 0$ such that $\A_x=c\A_y$. 
Further, it can be shown \cite{Kuramochi15b} that if $\A\sim\B$ are both simple, then they have the same number of outcomes and there is bijective relabeling that connects them. 
We denote by $\ell([\A])$ the number of POVM elements of the simple representative of $[\A]$. 

\begin{remark}
We conclude from the previous discussion that the equivalence classes are in one-to-one correspond to finite subsets of effects $\{ E_1, E_2, \ldots,E_n\}\subset\eh$ such that 
\begin{itemize}
\item[(i)] $\{ E_1, E_2, \ldots,E_n\}$ does not contain collinear operators;
\item[(ii)] $\sum_i E_i = \id$.
\end{itemize}
Hence, we can identify $\obsd^\sim$ with the collection of these type of subsets of $\eh$ whenever it is convenient.
\end{remark}

The structure of the partially ordered set $(\obsdsim, \pleq)$ is governed by the fact that the equivalence class of trivial POVMs( denoted by  $[\id]$)  is the least element and the maximal elements are those $[\A]$ with its simple representative $\A$ having $\operatorname{rank}(\A_i)=1$ for all $i \in \Omega_\A$ (some other representative consists also of rank-1 operators apart from zero effects); see Figure \ref{fig:obsdsim}, the left panel.
We further endow the set $\obsdsim$ with a convex structure as follows: for two equivalence classes $[\A]$ and $[\B]$ and $\lambda \in [0,1]$, we define
\begin{equation}\label{eq:convex-structure-traingle}
\lambda[\A] + (1-\lambda)[\B] := [\lambda \A \sqcup (1-\lambda) \B].
\end{equation}
In terms of the simple representatives, the convex combination of two equivalence classes is given by the class of the measurement obtained by taking the (weighted) \emph{union} of the effects from each class. 
We remark that $\lambda \A \sqcup (1-\lambda) \B$ is a different notion of convex mixture than the usual one for measurements, where one mixes corresponding effects of two measurements with equal number of outcomes. 
The definition above does not depend on the representatives chosen: assuming that $\A' \sim \A$ and $\B' \sim \B$, we observe that $\lambda \A \sqcup (1-\lambda)\B \sim \lambda \A' \sqcup (1-\lambda)\B'$ by taking direct sums of the respective column stochastic matrices. 

%%%%%%%%%%%%%%
\subsection{(In)compatibility of measurements}
%%%%%%%%%%%%%%

\begin{figure}
\centering
\includegraphics[width=8cm]{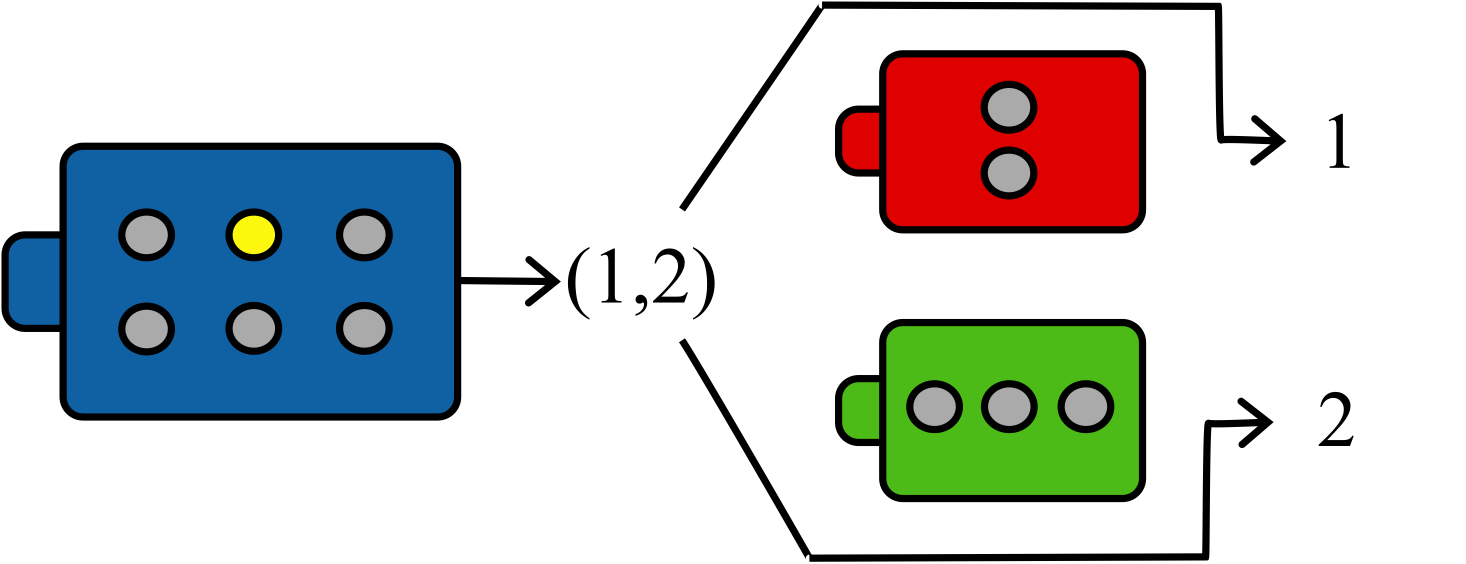}
\caption{In a joint measurement, one measurement device (blue) simulates two given measurement devices (red and green) by giving their outcomes as marginals. In this particular case the blue device gives the outcome $(1,2)$, from which one concludes that the outcome of the red device is $1$ and that of the green device is $2$.}
\label{fig:compatibility}
\end{figure}

As defined generally in Section \ref{sec:preorder}, a set of measurements $\mathcal{A} \subseteq \obsd$ is compatible if there exists a measurement $\C$ such that $\A\pleq\C$ for every $\A\in\mathcal{A}$; otherwise $\mathcal{A}$ is incompatible. 
In this context, compatibility has an interpretation as a joint measurement, also called simultaneous measurement. Equivalently (see \cite{AlCaHeTo09}), two measurements $\A$ and $\B$ are compatible if it exists a third measurement $\C$, with outcome set $X \times Y$, such that $\sum_y \C_{xy}=\A_x, \, \forall x$ and $ \sum_x \C_{xy}=\B_y, \, \forall y$. In this case, post-processing from $\C$ to $\A$ and $\B$ amounts to calculating the corresponding marginals; see Figure \ref{fig:compatibility}.

We can also introduce compatibility for equivalence classes of quantum measurements. 

\begin{definition}
A set $\mathcal{A}^{[]} \subseteq \obsdsim$ of equivalence classes of quantum measurements is \emph{compatible} if there exists a class $[\C]\in\obsdsim$ such that $[\A] \pleq [\C]$ for all $[\A] \in \mathcal A^{[]}$.
\end{definition}

This definition falls again into the general setting discussed in Section \ref{sec:preorder}: two equivalence classes of measurements $[\A]$ and $[\B]$ are compatible if there exists an element $[\C]$ in $(\obsdsim, \pleq)$ larger than both of them; see Figure \ref{fig:obsdsim}, right panel.
As compatibility is defined via post-processing, it is trivial that
$[\A]$ and $[\B]$ are compatible iff $\A$ and $\B$ are compatible. 

\begin{example}
Generally, commutativity is a sufficient condition for compatibility but not necessary.
It is a well-known fact that a quantum measurement $\A$ that consists of projections (i.e. $\A_x^2=\A_x$ $\forall x$) is compatible with another measurement $\B$ if and only if they commute, $\A_x \B_y = \B_y \A_x$ $\forall x,y$ (see e.g. \cite{HeReSt08}).  We conclude that the same is true for any measurement $\A' \in [\A]$. For instance, $\A'$ can have effects $\A'_1=a_1 \A_1$, $\A'_2=a_2 \A_1$ with $a_1 + a_2 =1$ and $0<a_1<1$. In this case $\A'_1$ is a not a projection but $\A'\in [\A]$ and the statement about the equivalence between commutativity and compatibility holds therefore for $\A'$.
\end{example}

%%%%%%%%%%%%%%%%%%%%%%%%%%%%%%%%%%%%%%%%%%%%%%%%
\section{Construction of order preserving maps}\label{sec:order-maps}
%%%%%%%%%%%%%%%%%%%%%%%%%%%%%%%%%%%%%%%%%%%

In this section, we introduce the notion of \emph{order preserving maps}, the main tool we shall use to study the post-processing partial order and the compatibility structure of quantum measurements. The main idea behind these considerations is that we want to map the partially order set of quantum measurements to a different poset, which has, simultaneously, the following two desirable properties: 
\begin{itemize}
    \item it is simpler, allowing for practical criteria for non-ordering and incompatibility
    \item it still captures enough of the structural properties of the quantum measurement poset to allow for powerful non-ordering and incompatibility criteria.
\end{itemize}
We will present several maps and make it precise how they can be compared.

%%%%%%%%%%%%%%%%%%%%%%%%%%%%%%%%%%%%%%%%%%%%%%%%%%%%
\subsection{General construction}
%%%%%%%%%%%%%%%%%%%%%%%%%%%%%%%%%%%%%%%%%%%%%%%%%%%

As mentioned already in Section \ref{sec:preorder}, we will study the partially ordered set $(\obsdsim, \pleq)$ by studying morphisms to simpler posets, such as the set of positive semidefinite matrices. 
In the following, we describe a general method to define order preserving maps from $(\obsdsim, \pleq)$ into a partially ordered vector space. Let $(\mathcal{V},\leq)$ be a partially ordered vector space, i.e.,~$\mathcal V$ is a real vector space, and $\leq$ is a relation on $\mathcal V$ satisfying the following two conditions
\begin{itemize}
    \item $x \leq y \implies x+z \leq y+z$ for all $x,y,z \in \mathcal V$
    \item $x \leq y \implies \lambda x \leq \lambda y$ for all $x,y \in \mathcal V$ and $\lambda \in \mathbb R$, $\lambda \geq 0$.
\end{itemize}
The set $\lhs$ of selfadjoint operators with the usual operator ordering is a paradigmatic example of a partially ordered vector space.
The following notion of order morphism is crucial for our investigation. 

\begin{definition}\label{def:order-morphism}
A map $g:\lhs \to \mathcal{V}$ such that
\begin{enumerate}[(i)]
\item $g(\lambda E)=\lambda g(E)$ for all scalars $\lambda \geq 0$.
\item $g(E+F) \leq g(E) + g(F)$
\end{enumerate}
is called an \emph{order morphism}. To any order morphism, we associate a map $G:\obsh \to \mathcal{V}$ given by
\begin{equation}\label{eq:G}
G(\A) = \sum_{x} g(\A_x) \, .
\end{equation}
\end{definition}

\begin{remark}
The property (ii) above corresponds to the definition of \emph{operator convex functions}, see \cite[Section 5]{MA97}. However, in that setting, the function $g(E)$ is defined using functional calculus (i.e.~$g$ is a real function which acts on the eigenvalues of the input), whereas we allow for more general matrix functions. In particular, asking that the property (i) above holds for a real function implies that the function is linear. 
\end{remark}

We note that in order for the map $G$ to be interesting and capture some of the order-related properties of the set $\obsh$, the map $g$ must be \emph{non-linear}. Indeed, if the map $g$ is linear, then we have 
    $$G(\A) = \sum_x g(\A_x) = g(\sum_x \A_x) = g(\id)$$
    and therefore $G$ is constant on $\obsh$.
    
The following is central for our investigation and links $G$ to our general discussion in Section \ref{sec:preorder} .

\begin{proposition}\label{prop:G-preserves-order}
The map $G$ defined in \eqref{eq:G} is order preserving on $\obsd$, i.e., $\A \pleq \B$ implies that $G(\A) \leq G(\B)$ for all $\A,\B\in\obsd$. 
Therefore, $G$ respects the equivalence relation of quantum measurements and it induces a map $G : \obsdsim \to \mathcal V$.
The induced map $G$ is order preserving on $\obsdsim$.
\end{proposition}

\begin{proof}
Let us first show the statement for $(\obsd, \pleq)$. Given two measurements $\A$, $\B$ with $\A \pleq \B$, there exists a column stochastic matrix $\mu$ such that \eqref{eq:pp-basic} holds. 
We then have
$$G(\A) = \sum_x g(\A_x) = \sum_x g\left(\sum_y \mu_{xy} \B_y\right) \leq \sum_x \sum_y \mu_{xy} g(\B_y) = \sum_y  \underbrace{\left(\sum_x \mu_{xy}\right)}_{=1}g(\B_y) = \sum_y g(\B_y) = G(\B).$$
It follows that $\A \pleq \B \pleq \A$ implies $G(\A)=G(\B)$, hence $G$ respects the equivalence relation. The last statement is a consequence of these two facts. 
\end{proof}

Proposition \ref{prop:f}(b) can be written in the following form in the present setting.

\begin{proposition}\label{prop:noelement}
Let $\mathcal{A}^{[]} \subset\obsdsim$.
If  there is no element $H\in G(\obsdsim)$ such that $G([\A])\leq H$ for all $[\A] \in \mathcal{A}^{[]}$,  then $\mathcal{A}$ is incompatible.
\end{proposition}

Deciding incompatibility of a collection of measurements by using Prop.~\ref{prop:noelement} requires that we know the range of $G$, or at least something about it. In this respect, an important fact is that he map $G$ is affine with respect to the convex structure of $\obsdsim$ introduced in \eqref{eq:convex-structure-traingle}.

\begin{proposition}\label{prop:G-respects-convexity}
	Given two measurements $\A$ and $\B$, as well as a convex weight $\lambda \in [0,1]$, the map $G$ respects the convex structure of the triangle introduced in \eqref{eq:convex-structure-traingle}, i.e., 
	$$G(\lambda[\A] + (1-\lambda)[\B]) = \lambda G([\A]) + (1-\lambda) G([\B]).$$
In particular, the range of the map $G$ is a convex set. 
\end{proposition}

\begin{proof}
The fact that $G$ is affine can be checked by using representatives of the classes $[\A], [\B]$:
	$$G(\lambda[\A] + (1-\lambda)[\B]) = \sum_{x=1}^{m} g(\lambda \A_x) + \sum_{y=1}^n g((1-\lambda)\B_y) = \lambda G([\A]) + (1-\lambda) G([\B]),$$
where we have used the fact that $\{\lambda \A_x\}_{x=1}^m \sqcup \{(1-\lambda)\B_y\}_{y=1}^n$ is a representative of the class $\lambda[\A] + (1-\lambda)[\B]$ defined in \eqref{eq:convex-structure-traingle}.
\end{proof}

%%%%%%%%%%%%%%%%%%%%%%%%%%%%%%%%%%%%%%%%%%%%%%%%%%%%%%%%%
\subsection{Quadratic maps}
%%%%%%%%%%%%%%%%%%%%%%%%%%%%%%%%%%%%%%%%%%%%%%%%%%%%%%%%

As noted before, linear maps $g$ are trivial. 
We focus now on a special class of order preserving maps, namely, that of \emph{quadratic maps}. 
These are maps $g$ of the form
\begin{equation}\label{eq:def-quadratic}
    g(E) = \frac{\Phi(E) \Phi(E)^*}{\tau(E)},
\end{equation}
where $\Phi : M_d(\mathbb C) \to M_{D \times r}(
\mathbb C)$ is a linear map and $\tau: M_d(\mathbb C) \to \mathbb C$ is a faithful positive linear form; in particular, we have $\tau(E) = \tr{\rho E}$ for some positive definite matrix $\rho$. (We can relax the last assumption on $\rho$ by requiring that the kernel of $\tau$ should be contained in the kernel of $\Phi$.) First, let us show that such maps verify the two axioms for order morphisms.

\begin{proposition}
The quadratic maps from Eq.~\eqref{eq:def-quadratic} satisfy the order morphism axioms of Definition~\ref{def:order-morphism}.
\end{proposition}
\begin{proof}
First, for a positive scalar $\lambda >0$, we have 
$$g(\lambda E) = \frac{\Phi(\lambda E) \Phi(\lambda E)^*}{\tau(\lambda E)} = \frac{\lambda^2\Phi(E) \Phi(E)^*}{\lambda\tau(E)} = \lambda g(E)$$
and hence the first axiom is satisfied.
Second, we have 
\begin{align*}
    \tau(E+F) & \left( g(E) + g(F) - g(E+F) \right)  \\
    & = \frac{\tau(F)}{\tau(E)} \Phi(E) \Phi(E)^* + \frac{\tau(E)}{\tau(F)} \Phi(F) \Phi(F)^* - \Phi(E) \Phi(F)^* - \Phi(F) \Phi(E)^* \\
    &= \left[ \sqrt{\frac{\tau(F)}{\tau(E)}}\Phi(E) - \sqrt{\frac{\tau(E)}{\tau(F)}} \Phi(F) \right] \left[ \sqrt{\frac{\tau(F)}{\tau(E)}}\Phi(E) - \sqrt{\frac{\tau(E)}{\tau(F)}} \Phi(F) \right]^* \geq 0 \, ,
\end{align*} 
showing that the second axiom is satisfied and concluding the proof.
\end{proof}

A standard choice for the form $\tau$ appearing in the denominator in \eqref{eq:def-quadratic} is $\tau(E) = \tr{E}$.
Regarding the choice of the linear map $\Phi$ in the numerator, several interesting choices can be made: 
\begin{enumerate}
	\item[(LIN1)] $\Phi(E) = \ket E \in \mathbb C^{d^2}$
	\item[(LIN2)] $\Phi(E) = E$
	\item[(LIN3)] $\Phi(E) = \operatorname{diag}(E)$
	\item[(LIN4)] $\Phi(E) = E\psi$ for a fixed unit vector $\psi\in\hi$
	\item[(LIN5)] $\Phi(E) = \tr{E}$.
\end{enumerate}

Here $\ket E$ denotes the vectorization of a matrix $E$, that is, the vector obtained by stacking the columns of $E$ on top of each other.
More exactly,  $$ \ket E=\sum\limits_{j,k=1}^d\bra k E \ket j \ket j \otimes \ket k\in \mathbb C^{d^2},$$ where $\{\ket j\}_{j=1}^d$ is an orthonormal basis on $ \mathbb C^{d}$.
 This map is an incarnation of the standard isomorphism between the vector space of linear maps between two vector spaces and their tensor product \cite[Section 10.2]{GQS06}. 
The first choice above corresponds to the \emph{Fisher information map} introduced by Huangjun Zhu \cite{Zhu15,ZhHaCh16} and will denote the respective quadratic map as $\fim$ and the derived order preserving map as $\Fim$.
This map will be discussed in detail in Section \ref{sec:FI-map}. 
Let us discuss, in reverse order, the other four examples above. In the following we use $\tau(E) = \tr{E}$ in the denominator if not otherwise stated.

\begin{itemize}

\item The map $\Phi(E) = \tr{E}$ is not a constant map, but it still gives a trivial map $G$ on $\obsd$ as $G(\A) = d$ for any measurement $\A$. It is hence useless for any purposes to study the order structure of $\obsdsim$ and we listed it to illustrate the fact that we cannot reduce the structure too much.

\item The choice $\Phi(E)=E\psi$, where $\psi \in \hi$ is a fixed unit vector, leads to the map
\begin{equation}\label{eq:G-psi}
G_\psi(\A) = \sum_x \frac{\A_x \ketbra{\psi}{\psi} \A_x^*}{\langle \psi |\A_x| \psi \rangle},
\end{equation}
where we have used the adapted choice $\tau(E) = \langle \psi| \A_x |\psi \rangle$. For the minimal class $[\id]$, we have $G_\psi([\id]) = \ketbra \psi \psi$, while for a rank-1 measurement $\A = (\A_x = \lambda_x\ketbra{a_x}{a_x})_x$, we have
$$G_\psi(\A) = \sum_x \lambda_x\frac{\ketbra{a_x}{a_x} \cdot \ketbra{\psi}{\psi} \cdot \ketbra{a_x}{a_x}}{|\langle a_x, \psi \rangle|^2} = \sum_x \lambda_x \ketbra{a_x}{a_x} = \id.$$

Hence, for an arbitrary measurement $\A$, we have $G_\psi(\A) \leq \id$. 
We conclude that this map cannot detect incompatibility. Namely, $G_\psi(\obsd)$ has the greatest element $\id$ and we are therefore in the situation of Example \ref{ex:catastrophe}.

\item For the choice $\Phi(E) = \operatorname{diag}(E)$ the map $G$ reads 
\begin{equation}\label{eq:G-4}
G(\A) = \sum_x \frac{\operatorname{diag}(\A_x)^2}{\tr{\A_x}}.
\end{equation}
Since, for diagonal matrices, the operator order reduces to coordinate-wise dominance, we obtain a necessary \emph{scalar} condition for the order, namely,
\begin{equation}\label{eq:scalar-condition}
[\A] \pleq [\B] \quad \Rightarrow \quad \forall k \in [d], \quad \sum_x \frac{\A_x(k,k)^2}{\sum_j \A_x(j,j)} \leq \sum_x \frac{\B_x(k,k)^2}{\sum_j \B_x(j,j)} \, .
\end{equation}
 As in the previous case, $G(\obsd)$ has the greatest element $\id$ hence the map $G$ cannot witness incompatibility of quantum measurements. However, the map $G$ can detect non-order in some particular cases. 

\item The choice $\Phi(E)=E$ gives $g(E)= E^2 / \tr{E}$. This yields the map
\begin{equation}\label{eq:G-3}
G(\A) = \sum_x \frac{\A_x^2}{\tr{\A_x}}.
\end{equation}
This map is easy to implement and, as we will shortly see, it is useful in investigating the order of $\obsdsim$.
On the downside, it cannot detect incompatibility.
Namely, for any rank-1 measurement $\A$ we have $G(\A) = \id$.
Since rank-1 measurements are maximal in $\obsd$, we have $G(\B)\leq \id$ for all $\B\in\obs$.
It follows that the image of $\obsdsim$ through this order morphism looks like a rhombus, with $\id$ on top (corresponding to rank-1 measurement) and $\id/d$ on the bottom (corresponding to trivial measurements).
This kind of image is useless in incompatibility detection as explained in Example \ref{ex:catastrophe}.

\end{itemize}

All of the introduced maps provide necessary conditions for two measurements being ordered. Namely, from Prop.~\ref{prop:G-preserves-order} we conclude that if the matrix $G([\A]) - G([\B])$ has both positive and negative eigenvalues, then neither $[\A] \pleq [\B]$ nor $[\B] \pleq [\A]$. This implication is only in one direction, but the test is easy to implement and therefore useful. For instance, let us consider the following two outcome, three outcome and four outcome qutrit measurements:
$$
\A_1= \frac{1}{6} \begin{pmatrix}
5 & 0 & 1 \\
0 & 4 & -1 \\
1 & -1 & 3
\end{pmatrix} \, , \quad 
\A_2= \frac{1}{6} \begin{pmatrix}
1 & 0 & -1 \\
0 & 2 & 1 \\
-1 & 1 & 3
\end{pmatrix} 
$$    
and
$$
\B_1= \frac{1}{12} \begin{pmatrix}
2 & 0 & 1 \\
0 & 1 & -1 \\
1 & -1 & 3
\end{pmatrix} \, , \quad 
\B_2= \frac{1}{12} \begin{pmatrix}
4 & -2 & 1 \\
-2 & 7 & 1 \\
1 & 1 & 7
\end{pmatrix} \, , \quad
\B_3= \frac{1}{6} \begin{pmatrix}
3 & 1 & -1 \\
1 & 2 & 0 \\
-1 & 0 & 1
\end{pmatrix}
$$    
and
$$
\C_1= \frac{1}{12} \begin{pmatrix}
7 & 0 & 2 \\
0 & 5 & -2 \\
2 & -2 & 6
\end{pmatrix} \, , \quad 
\C_2= \frac{1}{12} \begin{pmatrix}
1 & -1 & 0 \\
-1 & 3 & 1 \\
0 & 1 & 2
\end{pmatrix} \, ,
$$
$$
\C_3= \frac{1}{24} \begin{pmatrix}
5 & 2 & -1 \\
2 & 4 & 0 \\
-1 & 0 & 1
\end{pmatrix} \, , \quad
\C_4= \frac{1}{24} \begin{pmatrix}
3 & 0 & -3 \\
0 & 4 & 2 \\
-3 & 2 & 7
\end{pmatrix} \,.
$$    
Application of the map \eqref{eq:G-4} shows that both pairs $[\A], [\C]$  and  $[\B], [\C]$ are not ordered, but fails to indicate anything else. The map \eqref{eq:G-3}, on the other hand, shows that $[\A], [\B]$  and $[\B], [\C]$ are not ordered, but does not reveal anything else. Finally, using the Fisher information map $F$ shows that all three equivalence classes $[\A]$, $[\B]$ and $[\C]$ are pairwise not comparable.

%%%%%%%%%%%%%%%%%%%%%%%%%%%%%%%%%%%%%%%%%%%%%%%%%%%%%%%%
\subsection{Comparing order preserving maps}
%%%%%%%%%%%%%%%%%%%%%%%%%%%%%%%%%%%%%%%%%%%%%%%%%%%%%%%%

The difference between the quadratic maps that was noted in the end of last subsection motivates to compare order preserving maps in their power to detect non-orderings.
Clearly, if $[\A]\pleq [\B]$, then $G([\A]) \leq G([\B])$ for any order preserving map $G$. But, as shown in the previous subsection, there are also pairs $[\A],[\B]$, that are not comparable but still satisfy $G([\A]) \leq G([\B])$.
These are unwanted cases as then $G$ cannot observe the non-ordering.
The fewer of these cases we have,  the more informative is the order preserving map - this is the meaning of the following definition.

\begin{definition}\label{def:informative}
Let $G : \obsdsim \to \mathcal{V}$ and $G' : \obsdsim \to \mathcal{V}'$ be two order preserving maps.
We say that $G$ is at least as informative as $G'$ if for all quantum measurements $\A,\B$, the implication
\begin{equation}\label{eq:atleast}
G([\A]) \leq G([\B]) \quad \Rightarrow \quad G'([\A]) \leq G'([\B])
\end{equation}
holds.
We also say that $G$ is more informative than $G'$ if, in addition to \eqref{eq:atleast}, there exist $\A,\B$ such that $G'([\A]) \leq G'([\B])$ but  $G([\A]) \nleq G([\B])$.
\end{definition}

Our earlier example shows that the two quadratic maps given in \eqref{eq:G-4} and \eqref{eq:G-3} are not related in the sense of Def.~\ref{def:informative}, but that they are both more informative than the trivial map related to $\Phi(\A)=\tr{\A}$.
The content of the next result is that the Fisher information map is distinguished among quadratic order morphisms with $\tau(E) = \tr{E}$, in the sense that it is the most informative.

\begin{proposition}\label{prop:best}
	Among quadratic order preserving maps with $\tau(E)=\tr{E}$, the Fisher information map $\Fim$ is the most informative; it is at least as informative than any other order preserving quadratic map.
\end{proposition} 

\begin{proof}
A quadratic order morphism $g$ corresponds to a linear map $\Phi$ of the form 
$$\Phi(E) = \sum_{i=1}^k \langle \alpha_i, E\rangle \beta_i,$$ 
where $\alpha_i$ is a linear operator on $\mathbb{C}^d$, $\langle\cdot,\cdot\rangle$ is the Hilbert-Schmidt inner product ($\langle A,B\rangle=\langle A|B\rangle$) and $\beta_i\in M_{D \times r}$.
We thus have 
\begin{equation}\label{eq:Zhu-vs-quadratic}
    \Phi(E) \Phi(E)^* = J \ket{E} \bra{E} J^*
\end{equation}
for $J = \sum_{i=1}^k \beta_i \otimes \bra{\alpha_i}$. This shows that any such map is a post-processing (by conjugating with the $J$ matrix) of the Fisher information map.
Denoting by $\Fim$ the map that corresponds $\Phi(E) = \ket E$, we have from \eqref{eq:Zhu-vs-quadratic}, for any measurement $\A$,
\begin{equation}\label{eq:Zhu-best}
G(\A)=J \Fim(\A)J^* \, .
\end{equation}
Hence, $F(\A)\leq F(\B)$ implies $G(\A) \leq G(B)$.
\end{proof}

 We note that the proof of Prop.~\ref{prop:best} shows actually even more than is stated. Namely, by \eqref{eq:Zhu-best} we can get any other quadratic order preserving maps from the Fisher information map, assuming that we use the same map $\tau$ in the denominator.
 For instance, the matrices $J$ allowing to obtain the quadratic maps from Eqs.~\eqref{eq:G-4}, \eqref{eq:G-3} are given by:
    \begin{align*}
        J_{\eqref{eq:G-4}} &= \sum_{i=1}^d \ketbra{i}{iii}\\    
        J_{\eqref{eq:G-3}} &= \sum_{i,j=1}^d \ketbra{i}{ijj} = I_d \otimes \sum_{j=1}^d \bra{jj}.
    \end{align*}
    In the formulas above, the product $J \ket E$ is understood with the ket $\ket E$ acting on the first two Hilbert space factors. Moreover note that the map from Eq.~\eqref{eq:G-3} is the partial trace of the Fisher information map with respect to the second subsystem.
This conclusion naturally motivates to concentrate on the Fisher information map and that is what we do in the following sections.

%%%%%%%%%%%%%%%
\section{The generalized Fisher information map}\label{sec:FI-map}
%%%%%%%%%%%%%%%%

In \cite{Zhu15,ZhHaCh16} Huangjun Zhu and his collaborators considered the (classical) Fisher information matrix associated to a measurement $\A$ and to a statistical model centered at a given density matrix $\rho$. While in \cite{Zhu15} the focus was on the maximally mixed state $\rho = \id/d$, the case of a general quantum state $\rho$ was discussed at length in \cite{ZhHaCh16}. In this section, we view these constructions as order preserving maps from the set of measurements endowed with the post-processing partial order to the set of positive semidefinite matrices. We restate some of Zhu's results in the framework of Section \ref{sec:order-maps}, and we further develop the theory, adding new results and generalizations to this framework. 

%%%%%%%%%%%%%%%%
\subsection{Properties of the Fisher information map}
%%%%%%%%%%%%%%%

\begin{figure}
\centering
    \includegraphics{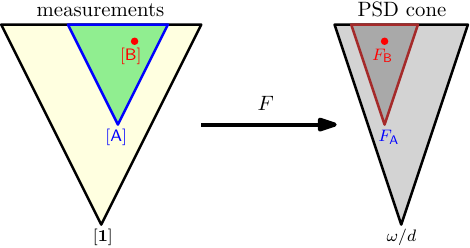}
    \caption{The Fisher information map $\Fim$ is an order morphism from the set of quantum measurements endowed with the post-processing order to the set of positive semidefinite matrices. We denote $F_X:=\Fim([X])$ for $X = \A,\B$.}
    \label{fig:thriangle-Zhu}
\end{figure}

Given $\rho$ a (fixed) faithful quantum state, that is $\rho > 0$, one can consider a generalized version of the Fisher information map induced by 
\begin{equation}\label{eq:def-Zhu-rho}
    \fimrho(E)=\frac{\ket{\rho^{1/2} E }\bra{  E \rho^{1/2}} }{\tr{\rho E}}.
\end{equation}
In the case of the (un-normalized) maximally mixed state, we simply write (see Figure \ref{fig:thriangle-Zhu} for a graphical representation):
\begin{equation}\label{eq:def-Zhu-original}
\fim(E)=\frac{\ketbra E E}{\tr{E}}.
\end{equation}
We denote by $\Fimrho$ and $\Fim$ the maps that correspond to $\fimrho$ and $\fim$, respectively, via \eqref{eq:G}.

Most of the following important results are contained in \cite{Zhu15,ZhHaCh16}. The new result here is the upper bound $\tr{\Fim([\A])} \leq \ell([\A])$, where $\ell([\A])$ is the number of elements in the simple representative of the equivalence class; see Sec.~\ref{sec:compatibility}.  For the sake of completeness, we prove all inequalities and also characterize the cases where the equality holds. 

\begin{proposition}\label{prop:zhu-map-properties}
	For any equivalence class $[\A] \in \obsdsim$ and for any given faithful state $\rho>0$, we have 
	$$ 1 \leq \tr{\Fimrho([\A])} \leq \min(d, \ell([\A]).$$
	The equality cases are as follows: 
	\begin{itemize}
		\item $\tr{\Fimrho([\A])} = 1$ iff $[\A]$ is the least element in $\obsdsim$, i.e., the class of trivial measurement. 
		\item $\tr{\Fimrho([\A])} = d$ iff $[\A]$ is a maximal element in $\obsdsim$, i.e., a class consisting of rank-1 measurements
		\item $\tr{\Fimrho([\A])} = \ell([\A])$ iff the simple representative of $[\A]$ is a sharp measurement, i.e., the operators in the range are projections.
	\end{itemize}
\end{proposition}

\begin{proof}
	For the first upper bound, consider a representative $\A \in [\A],$ with $ \A=(\A_x)_{x=1}^n$. The effects $\A_x$ are positive semidefinite as well as the faithful state $\rho>0$, hence, using the sub-multiplicativity of the trace, it holds that
	$\tr{\rho \A_x^2} \leq \tr{\rho \A_x}\tr{\A_x}$. 
	This implies
	$$\tr{\Fimrho(\A)} = \sum_{x=1}^{n} \frac{\tr{\rho \A_x^2}}{\tr{\rho \A_x}} \leq  \sum_{x=1}^n \frac{\tr{\rho \A_x}\tr{\A_x}}{\tr{\rho \A_x}} = \sum_{x=1}^n \tr{\A_x} = \tr{I_{\mathcal H}} =d \, .$$
	 The equality $\tr{\rho \A_x^2} = \tr{\rho \A_x}\tr{\A_x}$ holds if and only if $\A_x$ has rank one. 
	
	For the second upper bound, consider a simple representative $\A \in [\A]$ (see Subsec.~\ref{sec:simple}). Since $\A_x$ are effects, we have $\rho^{1/2} \A_x^2 \rho^{1/2}\leq \rho^{1/2} \A_x\rho^{1/2}$. Therefore, 
	$$\tr{\Fimrho(\A)} = \sum_{x=1}^{\ell(\mathcal A)} \frac{\tr{\rho \A_x^2}}{\tr{\rho \A_x}} \leq  \sum_{x=1}^{\ell(\mathcal A)} \frac{\tr{\rho \A_x}}{\tr{\rho \A_x}} = \ell(\mathcal A) \, .$$
	 The equality takes place if and only if for all $x$ and given $\rho> 0$, it holds  $\tr{\rho \A_x^2} = \tr{\rho \A_x}$ which is equivalent to the $\A_x$ being orthogonal projections.  
	
	Let us now turn to the lower bound. For a representative $\A \in [\A]$ and given $\rho>0$, by the Cauchy-Schwarz inequality, we have, for all $x$, $$
	\left(\tr{\rho \A_x}\right)^2=\left(\tr{\rho^{1/2}(\rho^{1/2} \A_x)}\right)^2 \leq \tr{\rho}\cdot \tr{(\rho^{1/2}\A_x)^*\rho^{1/2}\A_x}  =  \tr{\rho \A_x^2}\,. 
	$$
	This implies
	$$
	\tr{\Fimrho(\A)}= \sum_{x=1}^{n} \frac{\tr{\rho \A_x^2}}{\tr{\rho \A_x}} \geq  \sum_{x=1}^n \frac{\left(\tr{\rho \A_x}\right)^2}{\tr{\rho \A_x}} = \sum_{x=1}^n \tr{\rho \A_x} = 1 \, .
	$$
	Equality in the Cauchy-Schwarz inequality above holds iff each effect is a multiple of the identity, i.e.~the measurement $\A$ is trivial.	
\end{proof}

For $\rho=\id/d$, the inequality $\tr{\Fim(\A)} \leq d$ is the well-known Gill-Massar inequality \cite{GiMa00} for the $d$-dimensional Fisher information matrix. This was the original approach of \cite{Zhu15}. Let us also note that, in formula \eqref{eq:def-Zhu-rho}, one can have zero denominators if $\tr{\rho \A_x} = 0$. This can only occur if the positive semidefinite operators $\rho$ and $\A_x$ have orthogonal supports, in which case the numerator is also null. 
Hence, one can extend the framework by allowing non-invertible matrices $\rho$ simply by omitting terms with null denominators from the sum. For the sake of simplicity, we shall only consider invertible matrices $\rho$ in the current paper. 
We also note that using rank-one operator $\rho$ is useless for incompatibility detection. 
Namely, let us now consider a rank-one operator $\rho = \ketbra{\psi}{\psi}$, where $\psi \in \hi$ is a unit vector.  The generalized Fisher information map reads in this case $$F_{\ketbra \psi \psi}(\A) = \sum_x \frac{\Big | \ketbra \psi \psi \A_x \Big \rangle \Big \langle \ketbra \psi \psi \A_x \Big |}{\langle \psi | \A_x | \psi \rangle} = \ketbra \psi \psi \otimes \sum_x \frac{\A_x^\top \ketbra{\bar \psi}{\bar \psi} \overline{\A_x}}{\langle \psi | \A_x | \psi \rangle} = \ketbra \psi \psi \otimes G_{\bar \psi}(\A^\top),$$
where $G_{\bar \psi}$ is the map from Eq.~\eqref{eq:G-psi} and the complex conjugation is taken in the canonical basis. In particular, such a map cannot detect incompatibility, so the generalized Fisher information map with pure $\rho$ is useless as an incompatibility criterion.

We note that one can consider the ``truncated'' map
\begin{equation}\label{truncated}
    \bar{\Fimrho}(\A)=\sum\limits_x\frac{1}{\tr{\rho \A_x}}|\rho^{1/2}\bar{\A}_x\rangle \langle \bar{\A}_x\rho^{1/2}| \, , 
\end{equation}
where $\bar{\A}_x:=\A_x-\tr{\rho \A_x} \id$ is traceless with respect to $\rho$, i.e.~${\tr{\rho \bar{\A}_x}}=0$. 
It holds that
\begin{equation}\label{connection}
    \Fimrho(\A)-\bar{\Fimrho}(\A)= (\rho^{1/2} \otimes \id) \omega(\rho^{1/2} \otimes \id) \, ,
\end{equation}
where $\omega$ is the un-normalized maximally entangled density matrix $\omega = \ket \Omega \bra \Omega$ (or the identity supe-operator). Working with this version of the map can have the benefit that trivial measurements are mapped to the zero matrix. However, for the sake of simplicity, we shall focus in this work on the ``full'' version of Fisher information map from \eqref{eq:def-Zhu-rho} which is such that, for any trivial measurement $\T$, we have
	$$
	\Fimrho(\T) = (\rho^{1/2} \otimes \id) \omega(\rho^{1/2} \otimes \id) \, .
	$$

In the following we discuss two important properties of the Fisher information map; the first is related to combining of two measurements on separate systems while the second one is related to adding noise to a measurement.

\begin{proposition}
The map $\Fimrho$ respects tensor products: for any positive semidefinite matrices $\rho \in \M_d(\C)$, $\sigma \in \M_{d'}(\C)$, and measurements $\A,\B$ of appropriate dimensions, we have 
$$
F_{\rho \otimes \sigma}(\A \otimes\B) = F_\rho(\A) \otimes F_\sigma(\B),
$$
where $F_{\rho \otimes \sigma}(\A \otimes\B) :=\sum\limits_{x,y} \frac{\ket {\sqrt{\rho \otimes \sigma}\A_x \otimes\B_y}\bra{\A_x \otimes\B_y \sqrt{\rho \otimes \sigma}}}{\tr{(\rho \otimes \sigma)(\A_x \otimes\B_y)}}$.
\end{proposition}
\begin{proof}
The result follows from a simple computation, using the factorization property of the vectorization operation:
$$\ket{X \otimes Y} = \ket X \otimes \ket Y.$$
\end{proof}

A noisy measurement is typically described as a mixture of the original measurement with some trivial measurement \cite{DeFaKa19}. We remark that the mixture defined in the next statement is different than the one used in Prop.~\ref{prop:G-respects-convexity}, where the measurements are being concatenated.

\begin{proposition}
	For any measurement $\A$ and every $\lambda\in [0,1]$, we denote
	$$
	\A^{(\lambda)}_i := \lambda \A_i + (1-\lambda) \tr{\rho \A_i} \id \, .
	$$
	Then, for any given faithful state $\rho>0$, we have	
	\begin{equation}
	\Fimrho(\A^{(\lambda)})=\lambda^2 \Fimrho(\A)+(1-\lambda^2) \rho^{1/2} \omega \rho^{1/2} \, , 
	\end{equation}
	and
	\begin{equation}\label{eq:noise-scaling}
	\overline{\Fimrho}(\A^{(\lambda)})=\lambda^2 \overline{\Fimrho}(\A) \, .
	\end{equation}
	\end{proposition}
	
\begin{proof}
These are direct calculations.
\end{proof}

Especially \eqref{eq:noise-scaling} makes it effortless to study the effect of this kind of noise to measurements and their incompatibility. 

%%%%%%%%%%%%%
\subsection{Fisher information map on specific classes of measurements}
%%%%%%%%%%%%%

The Fisher information map has a particular form for sharp measurements, making it suitable for analysis.

\begin{proposition}
	If $\A$ is a sharp measurement, then both $\Fim(\A)$ and $\bar{\Fim}(\A)$ are orthogonal projections of respective ranks $\ell(\A)$ and $\ell(\A)-1$. 
\end{proposition}
\begin{proof}
The equality $\Fim(\A)^2 = \Fim(\A)$ follows by a direct computation. The maps $\Fim(\A)$ and $\bar{\Fim}(\A)$  are connected by \eqref{connection}. Therefore, it is straightforward to see that $\bar{\Fim}(\A)$ is also a projection and, by  \eqref{prop:zhu-map-properties}, has the rank one less than that of $\Fim(\A)$. 
\end{proof}

The Fisher information map is not injective, hence it does not preserve the full order structure of the triangle. However, in the following we show that it is one-to-one with respect to the subset of sharp measurements (i.e.~those corresponding to projective POVMs) and the subset of dichotomic measurements.
 \begin{proposition}\label{prop:injective-projective-measurement}
	Let $\A$ and $\B$ be two \emph{sharp} measurements such that $\Fim(\A) = \Fim(\B)$. Then $\A$ and $\B$ are post-processing equivalent: $[\A] = [\B]$.
\end{proposition}

\begin{proof}
	First, using the trace of the operator $\Fim(\A) = \Fim(\B)$ and Prop.~\ref{prop:zhu-map-properties}, we conclude that the measurements $\A,\B$ have the same number of outcomes. Let this be $k$. 
	Let $\{a_1, \ldots, a_d\}$ be an orthonormal basis of $\mathbb C^d$ in which the projections $\A_x$ are diagonal. By a direct computation, we have 
	$$\sum_{i=1}^d \bra{a_i \otimes \bar a_i} \Fim(\A) \ket{a_i \otimes \bar a_i} = k \, .$$
	We then have
	$$k = \sum_{i=1}^d \bra{a_i \otimes \bar a_i} \Fim(\B) \ket{a_i \otimes \bar a_i} = \sum_{i=1}^d \sum_{j=1}^k \frac{\bra{a_i} \B_j \ket{ a_i}^2}{\tr{\B_j}} \leq \sum_{i=1}^d \sum_{j=1}^k \frac{\bra{a_i} \B_j \ket{a_i}}{\tr{\B_j}} = \sum_{j=1}^k \frac{\tr{\B_j}}{\tr{\B_j}} = k.$$
	Since the equality holds above, we have that, for all $i,j$, $\bra{a_i} \B_j \ket{a_i} \in \{0,1\}$, showing that the projections $\B_j$ are also diagonal in the basis $\{a_1, \ldots, a_d\}$. In turn, this shows that the projections of $\A$ form a sub-partition of the projections of $\B$. We can then have the same argument by replacing the roles of $\A$ and $B$, hence the projections of $\B$ form a sub-partition of the projections of $\A$. Therefore, $\A$ and $\B$ are post-processing equivalent. 
\end{proof}

\begin{example}
	For a von Neumann measurement $\A$, defined as $\A_i =  \ket{e_i}\bra{e_i}$ for a basis $\{ e_i \}$, we have 
		$$\Fimrho(\A) = \sum_{i=1}^d \frac{[(\rho^{1/2}e_i )\otimes \bar e_i][e_i \otimes \overline{(e_i \rho^{1/2})}]^*}{ \tr{\rho e_ie_i^*}}= \sum_{i=1}^d \frac{\rho^{1/2}\ket{e_i} \bra{e_i} \otimes \overline{ \ket{e_i} \bra{e_i}}\rho^{1/2}}{ \tr{\rho \ket{e_i}\bra{e_i}}}.$$
\end{example}

We can also show that the Fisher information map is injective on dichotomic measurements. This is a simple consequence of the following expression: for a dichotomic measurement $\A$ consisting of effects $E$ and $\id-E$, we have
$$
\Fimrho(\A) =\rho^{1/2}\left[ \ketbra{\alpha \id + \beta E}{\alpha \id + \beta E} + \omega/ \tr{\rho}\right]\rho^{1/2},$$
where 
$$\alpha = \sqrt{\frac{\tr{ \rho E}} {\tr{\rho}\,\tr{\rho(\id-E)}}} \quad \text{and} \quad 
\beta = -\sqrt{\frac{\tr{\rho}}{\tr{\rho E}\tr{\rho(\id-E)}}}.$$
In the case where $\rho=\id,$ we get

$$\alpha = \sqrt{\frac{\tr{ E}}{d(d-\tr{E})}} \quad \text{and} \quad \beta = -\sqrt{\frac{d}{\tr{E}(d-\tr{E})}}.$$
Hence, one can recover the operator $E$ from the knowledge of ~$\Fimrho(\A)$: from the knowledge of $\rho$, $\Fimrho(\A)$ and the fact that  $\id$ and $E$ are independent (otherwise E is a trivial effect), one can recover $\ketbra E E$ and then, up to a phase, $E$.

We will then discuss an important class of measurements for which the Fisher information map is not injective.
We start by recalling that, as was pointed out in \cite{ZhEn11}, the non-singularity of the image of the map $\Fim$ is related to the measurement being informationally complete (i.e. the linear span of $\{\A_i\}_i$ is $\lh$). 

\begin{proposition}
Let $\rho > 0$. A measurement $\A$ is informationally complete iff the matrix $\Fimrho(\A)$ is invertible. 
\end{proposition}

\begin{proof}
It is a well-known fact (see e.g. \cite{Scott06}) that $\A$ is informationally complete iff the frame operator
$$
\mathcal{F}_\A = \sum_x \ket{\A_x}\bra{\A_x}
$$
is invertible. 
This happens iff the $d^2 \times d^2$ matrix $\Fimrho(\A)$ is invertible, since all the denominators $\tr{\rho \A_x}$ are positive. 
\end{proof}

For an arbitrary (i.e.~possibly continuous) measurement $\C$, we define the Fisher information map as follows: 
$$\Fim(\C) = \int_\Omega \frac{\ketbra{\C(\omega)}{\C(\omega)}}{\tr{\C(\omega)}} \, \mathrm{d}\omega.$$

\begin{example}
	For the continuous POVM $\C = d\{\ket \psi \bra \psi\}$, where $\psi$ is a uniformly distributed point on the unit sphere of $\mathbb C^d$, we have 
	$$\Fim(\mathsf C) = \frac{2}{d+1} P_{sym}^\Gamma,$$
	where $P_{sym}$ is the orthogonal projection on the symmetric subspace of $\mathbb C^d \otimes \mathbb C^d$, and the $\Gamma$ symbol denotes the partial transposition. Indeed, using $\big | \ketbra{\psi}{\psi} \big\rangle = \ket \psi \otimes \overline{\ket \psi}$, we have 
	\begin{align}\label{eq:continuous-measurement}
	\Fim(\mathsf C) & = d\int_{\|\psi\| = 1} \ketbra{\psi}{\psi} \otimes \overline{\ketbra{\psi}{\psi}} \, \mathrm{d} \psi = d \left[ \int_{\|\psi\| = 1} \ketbra{\psi}{\psi} \otimes {\ketbra{\psi}{\psi}} \, \mathrm{d} \psi \right]^\Gamma \\
	& = d \frac{P_{sym}^\Gamma}{d(d+1)/2} = \frac{2}{d+1} P_{sym}^\Gamma.
	\end{align}
	The third equality above follows from \cite[Proposition 2]{harrowsym}. Note that above, $\mathrm{d}\psi$ denotes the \emph{normalized} Lebesgue measure on the unit sphere of $\mathbb C^d$, in such a way that 
	$$\int \ketbra \psi \psi \, \mathrm{d}\psi = \frac{\id}{d}.$$
\end{example}

\begin{proposition}
	Let $\mu$ be a complex spherical 2-design, that is $\mu$ is a probability measure on the unit sphere of $\mathbb C^d$ such that 
	\begin{equation}\label{eq:2-design}
	\binom{n+d-1}{n} \int \ket \psi \bra \psi ^{\otimes n} \mathrm{d}\mu(\psi) = P^{(n)}_{sym} \qquad \text{ for } n=1,2,
	\end{equation}
	where $P^{(n)}_{sym}$ is the orthogonal projection on the symmetric subspace of $(\mathbb C^d)^{\otimes n}$. Then, the POVM $\A = d\{ \mu(\psi) \ket \psi \bra \psi \}$ is such that
	$$\Fim(\A) = \frac{2}{d+1} [P_{sym}^{(2)}]^\Gamma.$$
\end{proposition}
\begin{proof}
	First, note that \eqref{eq:2-design} for $n=1$ is precisely the condition that $\A$ is a POVM. For $n=2$, equation \eqref{eq:continuous-measurement} still holds when replacing the Lebesgue measure $\mathrm{d} \psi$ by the 2-design $\mathrm{d}\mu(\psi)$.
\end{proof}

\begin{corollary}
	The Fisher information map map $\Fim$ is not injective, since any 2-design is mapped to $2P_{sym}/(d+1)$. 
\end{corollary}

\begin{remark}
	Let $\A$ be a rank-1 measurement, i.e., a maximal element $\obsdsim$. Then
	$$\Fim(\A)^\Gamma \in \mathrm{conv}\{\ket a \bra a^{\otimes 2} \, : \, a \in \mathbb C^d\}.$$
	In particular, $\Fim(\A)^\Gamma$ is a separable, symmetric, positive semidefinite matrix. 
\end{remark}

%%%%%%%%%%%%%%%
\subsection{Measurement incompatibility criterion}\label{sec:incompatibility-criterion}
%%%%%%%%%%%%%%%%

Up to this point, we have focused on the properties of the Fisher information maps relative to the partial order of post-processing. The paramount application of this map is to detect incompatibility of measurements and this is the topic of the current section. We gather here the main theoretical results and we present applications in the next section. 

The incompatibility criterion for quantum measurements that we obtain can be expressed with the help of the following height function for matrices. 

\begin{definition}
For a finite set of $d \times d$ self-adjoint matrices $X = \{X_i\}_{i \in [n]}$, define the \emph{height} of $X$ as
\begin{equation}\label{eq:def-h}
\mathfrak h(X) := \min\{ \tr{H} \, : \, H \geq X_i \quad \forall i \in [n]\}.
\end{equation}
\end{definition}

\begin{figure}
    \centering
    \includegraphics[scale=1]{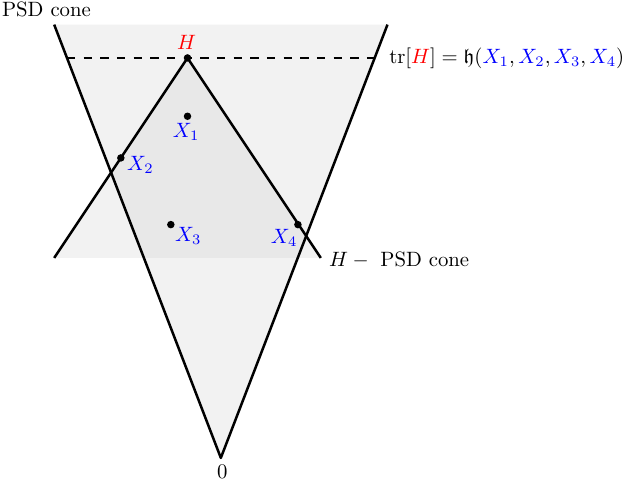}
    \caption{The height function of a set of four matrices $\textcolor{blue}{X_{1,2,3,4}}$. Note that in general, the set $\{\textcolor{blue}{X_i}\}$ does not have a smallest upper bound in the PSD order. }
    \label{fig:height}
\end{figure}

In Figure \ref{fig:height} we provide an interpretation of the height function: it is the minimum trace of an operator $H$ which dominates, in the PSD order, the set $\{X_1, \ldots, X_n\}$. Since the positive semidefinite (or L\"owner) cone is not a lattice, the set of matrices $H$ above has a complicated structure (e.g.~it does not admit a least element in general), and thus the minimization problem is not trivial. In some cases height function is easy to calculate. For instance, if the matrices in $X$ are commuting, they can be diagonalized in some common basis, and one can easily find the smallest (entry-wise) diagonal matrix $H$ dominating all the $X_i$.

A crucial fact about the height function is that it can be computed with the help of a semidefinite program (SDP) \cite{CO04}. In the case of two matrices, one can get an explicit formula for $\mathfrak h$, see \cite[Eqs.~(13)-(15)]{HeJiNe20}:
\begin{equation}\label{SDP}
\mathfrak h(X_1,X_2) = \frac 1 2 \left( \tr{X_1} + \tr{X_2} + \|X_1 - X_2\|_1 \right),\end{equation}
where $\| \cdot \|_1$ denotes the Schatten $1$-norm (or the nuclear norm).

In the general setting, the height function satisfies the following set of properties. 

\begin{proposition}\label{lem:orthogonal-supports}
Given positive semidefinite matrices $X_1, \ldots, X_n$, the following upper bound holds:
$$\mathfrak h(X) \leq \sum_{i=1}^n \tr{X_i}.$$
Moreover, if the matrices $X_i$ have orthogonal supports, the inequality above is saturated.
\end{proposition}

\begin{proof}
The first claim follows from the fact that $H = \sum_{i=1}^n X_i$ is feasible point of the semidefinite program \eqref{eq:def-h} and hence, $ \tr{H} \geq  \tr{ \sum_{i=1}^n X_i}$. The second claim follows from the following implication, which is true if the matrices $X_i$ have orthogonal supports: 
$$\forall i \in [n], \quad H \geq X_i \implies H \geq \sum_{i=1}^n X_i.$$
\end{proof}

One can compute the dual of the SDP in Eq.~\eqref{eq:def-h} \cite[Chapter 5]{CO04} and obtain: 
\begin{equation}\label{eq:Zhu-dual}
\mathfrak h(X) = \max \left\{\sum_{i=1}^n \tr{X_iY_i} \, : \, Y  = (Y_1, \ldots, Y_n) \text{ is a $n$-outcome measurement}\right\}.
\end{equation}

In particular, any choice of a measurement $Y$ gives a \emph{lower bound} on the quantity $\mathfrak h(X)$. Moreover, in the case when the matrices $X_i$ are positive semidefinite, an interesting choice is the \emph{pretty-good-measurement} \cite{HaWo94}:
$$Y_i = S^{-1/2} X_i S^{-1/2}, \qquad \text{ with } S = \sum_{i=1}^n X_i,$$
where one uses the pseudo-inverse to compute the inverse matrix square roots. 

The following result is the \emph{incompatibility criterion} discovered by Zhu \cite{Zhu15}. We slightly improve it below, by adding another constraint depending on the number of measurement outcomes.

\begin{theorem}\label{thm:Zhu-criterion}
Let $\mathcal{A}=\{ \A^{(1)}, \ldots, \A^{(g)}\}$ be a set of $g$ $d$-dimensional measurements. 
For any faithful quantum state $\rho$, we denote
$$
\Fimrho(\mathcal{A}):=\left\{\Fimrho(\A^{(i)})\right\}_{i=1}^g.
$$
If there exists $\rho>0$ such that
$$
\mathfrak h\left( \Fimrho(\mathcal{A}) \right) > \min\left(d, \prod_{i=1}^g \ell(\A^{(i)})\right)
$$
then $\mathcal{A}$ is incompatible.
\end{theorem}

\begin{proof}
Assuming that the given measurements are compatible, let $\B$ be their joint measurement. Since every $\A^{(i)}$ is a post-processing of $\B$, it follows from Prop.~\ref{prop:G-preserves-order} that $\Fimrho(\A^{(i)}) \leq \Fimrho(\B)$ for all $i \in [g]$ and thus, using \eqref{eq:def-h}, we get $\mathfrak h(\Fimrho(\mathcal{A})) \leq \tr{\Fimrho(\B)}$. 
From Prop.~\ref{prop:zhu-map-properties} we know that 
$$
\tr{\Fimrho(\B)} \leq \min(d,\ell(\B)) \, 
$$
and the conclusion follows from the fact that one can always choose a joint measurement having $\prod_{i=1}^g \ell(\A^{(i)})$ outcomes.
\end{proof}

\begin{remark}
If the measurements $\A^{(1)}, \ldots, \A^{(g)}$ are such that 
$$\sum_{i=1}^g\tr{F_\rho(\A^{(i)})} \leq \min\left(d, \prod_{i=1}^g \ell(\A^{(i)})\right),$$
it follows from Proposition \ref{lem:orthogonal-supports} that one cannot witness the incompatibility of these measurements using the criterion from Theorem \ref{thm:Zhu-criterion}.
\end{remark}

Using the dual formulation of the SDP used for defining the quantity $\mathfrak h$ and the pretty-good-measurement of \cite{HaWo94}, we obtain the following simpler incompatibility criterion, which does not require computing the value of a semidefinite program. 

\begin{corollary}
Let $\mathcal{A}=\{ \A^{(1)}, \ldots, \A^{(g)}\}$ be a set of $g$ $d$-dimensional measurements. 
We define a measurement $\Y$ with $g$ outcomes by
$$
\Y_i = S^{-1/2} \Fimrho(\A^{(i)}) S^{-1/2}, \qquad \text{ with } S = \sum_{i=1}^g \Fimrho(\A^{(i)}).$$
If 
$$
\sum_{i=1}^g \tr{\Fimrho(\A^{(i)}) \Y_i} > \min\left(d, \prod_{i=1}^g \ell(\A^{(i)})\right) \, , 
$$
then $\mathcal{A}$ is incompatible.
\end{corollary}

\begin{remark}
In the dual optimization problem from eq.~\eqref{eq:Zhu-dual}, one can recognize an expression similar to the one of the optimal zero-error, one-shot, distinguishability of $n$ quantum states of dimension $d^2$. 
Indeed, the function $\mathfrak h$ is a measure of distinguishability of the matrices $X_i$. Hence, the more distinguishable the matrices $X_i$ are, the more incompatible they are (in the sense of the criterion from Theorem \ref{thm:Zhu-criterion}).
\end{remark}

One can use the newly introduced bound from Proposition \ref{prop:zhu-map-properties} to give a lower bound on the number of outcomes of any joint measurement for a $g$-tuple of compatible measurements. 

\begin{proposition}
Let $\mathcal{A}=\{ \A^{(1)}, \ldots, \A^{(g)}\}$.
For any faithful quantum state $\rho>0$, the length of any joint measurement $\B$ (i.e. $\B \pgeq \A^{(i)}$ for all $i$) satisfies
$$
\ell(\B) \geq \mathfrak h(\Fimrho(\mathcal{A})) \, .
$$
\end{proposition}
\begin{proof}
The proof is a simple consequence of Proposition \ref{prop:G-preserves-order} and  Proposition \ref{prop:zhu-map-properties} (or directly from Theorem \ref{thm:Zhu-criterion} ).
\end{proof}

\begin{remark}
We recall that it has been shown in \cite{SkHoSaLi20} that, given a $g$-tuple of compatible measurements, one can always find a joint measurement with a number of outcomes which is \emph{linear} in the number $g$ of measurements. 
This is an interesting result, but it does not yield a better incompatibility criterion in Theorem \ref{thm:Zhu-criterion} since the value is always larger than $d$.
\end{remark}

%%%%%%%%%%%%%%%%%%%%%%
\section{Examples}\label{sec:examples}
%%%%%%%%%%%%%%%%%%%%%%%

In this section we investigate the efficiency of our approach in some concrete examples, for which we can find the incompatibility conditions for the given measurements and compare them to known necessary and sufficient conditions.
As we will demonstrate, the Fisher information incompatibility condition is exact in some cases, good in some cases, while relatively poor in some cases.

%%%%%%%%%%%%
\subsection{Fourier-conjugate bases}
%%%%%%%%%%%%

The compatibility problem for noisy versions of Fourier-conjugate bases has been studied previously in \cite{CaHeTo12,CaHeTo19}. 
We recall that two $d$-outcome measurements $\A$ and $\B$, are called the noisy version of Fourier-conjugate measurements if
\begin{align*}
\A_i &= s \ket{e_i}\bra{e_i} + (1-s)\frac{\id}{d}, \qquad s\in[0,1]\\
\B_j &= t \Fou \ket{e_j}\bra{e_j} \Fou^* + (1-t)\frac{\id}{d}, \qquad t\in[0,1]
\end{align*}
where $\Fou$ is the Fourier matrix $\Fou_{ab} = \zeta^{ab}/\sqrt d$ with $\zeta = \exp(2\pi \mathrm i / d)$ a primitive root of unity. 

\begin{figure}
\centering
\includegraphics[width=0.3\linewidth]{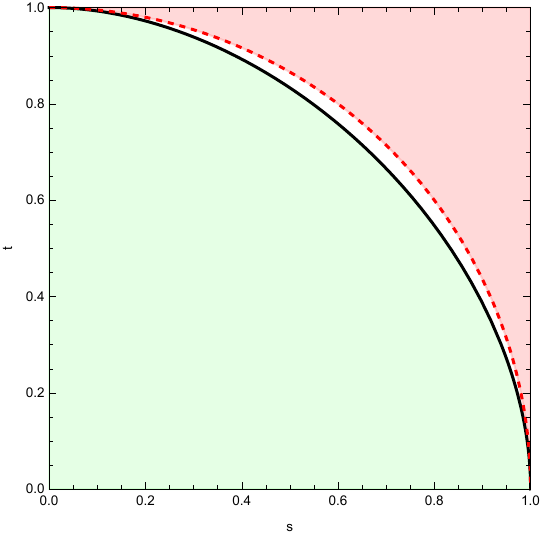} \quad
\includegraphics[width=0.3\linewidth]{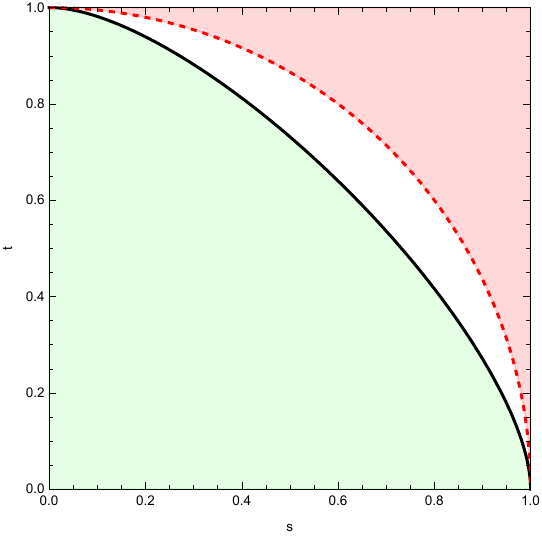} \quad
\includegraphics[width=0.3\linewidth]{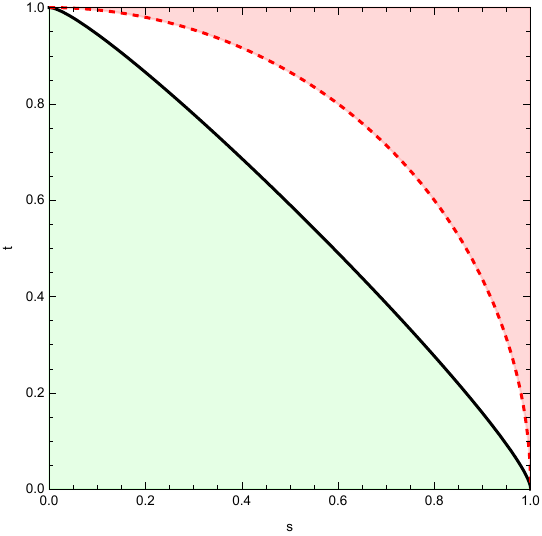}
\caption{The region of parameters $(s,t) \in [0,1]^2$ for which noisy versions of Fourier-conjugated bases are compatible: left $d=3$, center $d=10$, right $d=100$. The compatibility region computed in \cite{CaHeTo12} is delimited by the thick black curve, with the green shaded region corresponding to pairs $(s,t)$ of compatible measurements. The Fisher information incompatibility criterion corresponds to the thick dotted red curve. The red shaded region corresponds to pairs $(s,t)$ for which incompatibility is guaranteed by Zhu's Fisher information criterion. For $d=2$ the two curves overlap.}
\label{fig:fourier-conjugated}
\end{figure}

It has been shown in \cite[Prop. 6]{CaHeTo12} that $\A$ and $\B$ are compatible iff 
\begin{equation}
\label{reg-comp}s+t \leq 1\quad  \text{ or } \quad s^2+t^2 + \frac{2(d-2)}{d}(1-s)(1-t) \leq 1.
\end{equation}
Let us apply Fisher information incompatibility criterion from Section \ref{sec:incompatibility-criterion} in this situation. 
As it has been shown in \cite[Equation (23)]{Zhu15}, a necessary condition for compatibility is $s^2 + t^2 \leq 1$. 
The application of the criterion is straightforward, since for mutually unbiased bases (MUBs) $\A$ and $\B$, the matrices $\bar \Fim(\A)$ and $\bar \Fim(\B)$ are orthogonal, and Prop.~\ref{lem:orthogonal-supports} applies. We notice that for  $d=2$ the two conditions of incompatibility, that are the conditions  given by Theorem \ref{thm:Zhu-criterion} and equation \eqref{reg-comp}, are the same.  It is of interest to compare the  corresponding incompatibility regions of the parameters $(s,t)$  for different dimension values and this is presented in Figure \ref{fig:fourier-conjugated}. As we can see, in general, the shaded region that corresponds to pairs $(s,t)$ for which incompatibility is guaranteed by Zhu's Fisher information criterion is much smaller respect to the region given by \eqref{reg-comp}.

%%%%%%%%%%%%%
\subsection{Two unbiased dichotomic qubit measurements}
%%%%%%%%%%%%%%

We consider now the important examples of  unbiased dichotomic qubit measurements  given by $\A^{(k)}(\pm1)=\frac{1}{2}(\id\pm\eta_k\vec{n}_k\cdot \sigma),\,  k=1,2,3$, where $0\leq \eta_k\leq 1$ is the  sharpness, $\vec{n}_k=\vec{e_k},\,  k=1,2,3$ are the vectors of canonical basis in $\mathbb{R}^3$ and $\sigma=(\sigma_x,\sigma_y,\sigma_z)$ is the vector of the Paulli operators.
It has been shown in \cite{Busch86} that any two among these three measurements are compatible iff $\eta_k^2 + \eta_\ell^2 \leq 1$.

We aim to  compute in this case the Fisher information matrix \begin{equation*}\Fim_{\rho}(\A^{(k)}):=\sum_i \frac{\ketbra{\rho^{1/2}A^{(k)}_i}{\rho^{1/2}A^{(k)}_i}}{\tr{\rho A^{(k)}_i}},\end{equation*} for  $\rho=\frac{1}{2} \left(\begin{array}{cc}1+v_3&v_1-iv_2\\v_1+iv_2&1-v_3
\end{array}\right)$ and $v=(v_1,v_2,v_3)^T,||v||\leq1$ is the Bloch vector of the state $\rho$.

To exemplify our theory we present the  problem in the setting $v_1=v_2=0$.
In this  case, the Fisher information  matrices of the measurements $\A^{(k)},k=1,2,3$ can be written explicitly as 
\begin{eqnarray}\Fim_{\rho}(\A^{(1)})&=&\frac{1}{2}  \left(\begin{array}{cccc}\nonumber  (1+v_3)&0&0&\sqrt{1-v_3^2}\\0&\eta_1^2(1-v_3)&\eta_1^2\sqrt{1-v_3^2}&0
\\0&\eta_1^2\sqrt{1-v_3^2}&\eta_1^2(1+v_3)&0\\ \sqrt{1-v_3^2}&0&0&(1-v_3)\end{array}\right)\,\,\\
\Fim_{\rho}(\A^{(2)})&=&\frac{1}{2}  \left(\begin{array}{cccc}\nonumber  (1+v_3)&0&0&\sqrt{1-v_3^2}\\0&\eta_2^2(1-v_3)&-\eta_2^2\sqrt{1-v_3^2}&0
\\0&-\eta_2^2\sqrt{1-v_3^2}&\eta_2^2(1+v_3)&0\\ \sqrt{1-v_3^2}&0&0&(1-v_3)\end{array}\right) \,\,\\
\Fim_{\rho}(\A^{(3)})&=&\frac{1}{2}  \left(\begin{array}{cccc} (1+v_3)(a^2+1)&0&0&\sqrt{1-v_3^2}(1-ab)\\0&0&0&0
\\0&0&0&0\\ \sqrt{1-v_3^2}(1-ab)&0&0&(1-v_3)(b^2+1)\end{array}\right)  \end{eqnarray}
where 
$a=\frac{\eta_3-\eta_3v_3}{\sqrt{1-\eta_3^2v_3^2}},\, b=\frac{\eta_3+\eta_3v_3}{\sqrt{1-\eta_3^2v_3^2}}$.

It follows that 
 $\tr {\Fim_{\rho}(\A^{(1)})}=\eta_1^2+1$,  $\tr {\Fim_{\rho}(\A^{(2)})}=\eta_2^2+1+v_3$ and
$\tr {\Fim_{\rho}(\A^{(3}))}=1+\frac{\eta_3^2-\eta_3^2v_3^2}{1-\eta_1^2v_3^2}.$ 
We aim to compute the height of Fisher information matrices $\Fim_{\rho}(\A^{(k)}),k=1,2,3$, for finding the incompatibility region given by the Theorem \ref{thm:Zhu-criterion}.
In the case of two measurements, for example, $\A^{(1)}$ and $\A^{(3)}$ (corresponding to $\sigma_x$ and $\sigma_z$) we find that 
the incompatibility condition given by the SDP \eqref{SDP} is
\begin{equation}
\eta_1^2+\eta_3^2-\eta_1^2\eta_3^2v_3^2> 1, 
\end{equation} 
where $(\eta_1,\eta_3)\in[0,1]^2$.

We notice that for general $v_3 (|v_3|\leq 1)$ we get a region of compatibility  which is larger than the quatercircle  $\eta_1^2+\eta_3^2\leq 1$, corresponding to the case $v_3=0$ $ (\rho=\frac{\id}{2})$. For $v_3=\pm1$ ( corresponding to $\rho$ pure state) we get the full square, so the criterion is uninformative.
For $(\eta_1,\eta_3)=(1,0),$ we get that the compatibility holds for any $v_3,|v_3|\leq 1$.

In the case of the two sharp measurements corresponding to the observables $\sigma_x$ and $\sigma_y$, for any $v_3$, we get the incompatibility condition $\eta^2_1+\eta^2_2>1$. This result suggests that considering a state $\rho$ given by a Bloch vector of the form $v=(0,0, v_3)$ gives the same incompatibility condition, independent of $v_3$.

It is of interest to study the influence of a generic state $\rho$ in determining the incompatibility region for two unbiased dichotomic qubit measurements, for example, the  cases corresponding to $\sigma_x,$ $\sigma_z$ and  $\sigma_x,$  $\sigma_y$.
We are going to study this by computing the height function
$\mathfrak h\left( \Fimrho(\mathcal{A}) \right)\equiv
h(v_1,v_2,v_3,\eta_1,\eta_2,\eta_3)$ which is now a function of the sharpness $\eta_i, i=1,2,3$ and of the parameters of the Bloch vector $v_i,i=1,2,3$.
\begin{figure}
\centering
\includegraphics[width=0.45\linewidth]{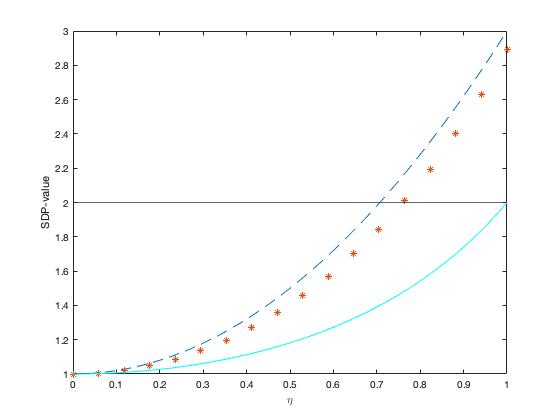}\quad
\includegraphics[width=0.45\linewidth]{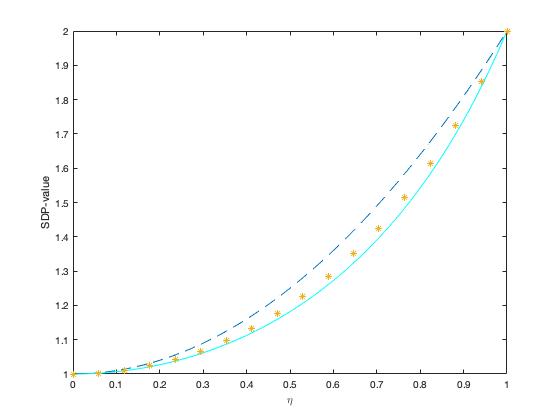} \quad
\includegraphics[width=0.5\linewidth]{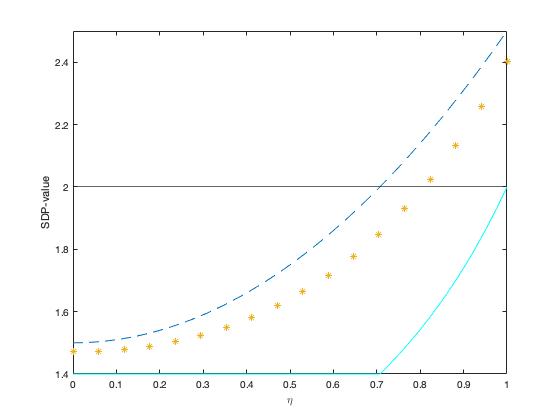}
\caption{The plot of the height function $h(v_1,v_2,v_3,\eta,\eta,\eta)$ (first),
 $h(v_1,v_2,v_3,0,\eta,0)$ (second)  and 
 $h(v_1,v_2,v_3,0,\eta,1/\sqrt{2})$ (third ) of the
     generalized Fisher information matrix corresponding to two unbiased dichotomic qubit measurements ( $\sigma_x,$ $\sigma_y$) for  $\rho=\frac{1}{2}(\id + v\cdot \sigma)$; $v=(0,0,0)$ is dashed ($--$) line, $v=(1/2,1/3,1/3,0)$ is star ($*$) line and $v=(1/\sqrt{3},1/\sqrt{3},1/\sqrt{3})$ for cyan line.}
\label{fig:XY1}
\end{figure}
\eqref{fig:XY1} 

\begin{figure}
\centering
\includegraphics[width=0.6\linewidth]{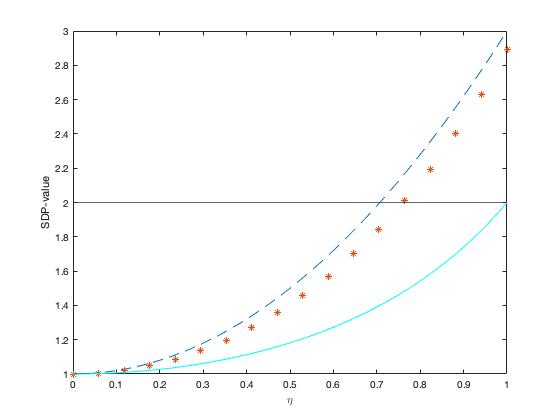} \quad
\caption{The plot of the height function $h(v_1,v_2,v_3,\eta,\eta,\eta)$ of the
     generalized Fisher information matrix corresponding to two unbiased dichotomic qubit measurements ( $\sigma_x,$ $\sigma_z$) for  $\rho=\frac{1}{2}(\id + v\cdot \sigma)$; $v=(0,0,0)$ is dashed ($--$) line, $v=(1/2,1/3,1/3)$ is star ($*$) line and $v(1/\sqrt{3},1/\sqrt{3},1/\sqrt{3})$ for cyan line.}
\label{fig:XZ}
\end{figure}

 As we can see from Figures \ref{fig:XY1} (the case $\sigma_x$ and $\sigma_y$) and \ref{fig:XZ} (the case $\sigma_x$ and $\sigma_z$), the limiting condition for incompatibility $h(v_1,v_2,v_3,\eta,\eta,\eta)> 2$ is better achieved using the Bloch vector $v=(0,0,0)$, that corresponds to the maximally mixed state, comparing to another specific Bloch vector, for example $v=(1/2,1/3,1/3)$. We also notice that using unit Bloch vector does not provide information about the region of incompatibility. Our graphics confirms the results that two unbiased dichotomic qubit measurements of the same sharpness $\eta$ are incompatible for $\eta>1/\sqrt{2}\approx 0.7$, which actually coincide with the value of $h(v_1,v_2,v_3,\eta,\eta,\eta)> 2$, for which it becomes grater that 2.

%%%%%%%%%%%%%%%%%%%
\subsection{Unbiased quantum effects}
%%%%%%%%%%%%%%%%%%%

Let $\A^{(i)}_\pm = ((\id \pm \T_i)/2)$ be a $g$-tuple of unbiased dichotomic measurements, for some traceless self-adjoint matrices $T_1, \ldots, T_g \in M_d(\mathbb C)$. A simple computation shows that the truncated map $\bar \Fim_i $ 
introduced in \eqref{truncated} can be written as
$\bar \Fim_i = \ket{\T_i}\bra{\T_i}/d$. Assume now that all the $\T_i$ are unitary and anti-commuting; such a family exists, for example, if $d\geq 2^{g-1}$, see \cite{Newman32} or \cite[Theorem 1]{Hrubes16}. It has been shown in \cite[Theorems VII.7 and VIII.8]{BlNe18} that such a $g$-tuple of quantum effects is maximally incompatible: the set of noise parameters $s \in [0,1]^g$ rendering the noisy measurements
$$\tilde \A^{(i)} = s_i \A^{(i)} + (1-s_i) (\id/2, \id/2)$$
compatible is precisely the ``quarter circle'' $\{s \in [0,1]^g \, : \, \sum_{i=1}^g s_i^2 \leq 1 \}$. The term \emph{maximal incompatibility} refers to the fact that noise parameters inside the quarter circle generate compatible dichotomic measurements, for \emph{any} starting measurements \cite{BuHeScSt13}. 
We analyze the next the power of Zhu's criterion in this setting. For unitary, self-adjoint, anti-commuting $T_i$'s, we have 
$$\tr{\T_i\T_j} = d \delta_{ij},$$
which implies that the matrices $\bar \Fim_i$ are $g$ orthonormal unit-rank projections. Zhu's criterion for the noisy versions reads
$$\sum_{i=1}^g s_i^2  \leq d-1,$$
which is precisely the quarter circle condition in the case $d=2$, and a much worse condition for larger $d$. For $d=2$, this condition proves the optimality of $g=2$ or $g=3$ unbiased qubit effects; the matrices $F$ are in this case the Pauli operators. The next cases, $g=4,5$, require $d=4$ elements, the condition above still being non-trivial. However, for $g \geq 6$, we require $d \geq 8$, yielding the condition trivial, since 
$$g \leq 2^{\lceil(g-1)/2 \rceil}-1.$$

%%%%%%%%%%%%%%%
\subsection{Planar qubit measurements}\label{subsec:planar-obs}
%%%%%%%%%%%%%%%

For the planar qubit effects defined in \cite{UoLuMoHe16}, after adding noise, the Zhu maps read
$$\bar \Fim(a) = \lambda^2 \ket a \bra a.$$
Requiring $H \geq \bar \Fim(a)$ for $a = \cos(k\pi /M) e_1 + \sin(k \pi / M) e_2$ for $k = 0, 1, \ldots, M-1$, the best choice is $H = \lambda^2 I_2$, which gives the inequality $\lambda \leq 1 /\sqrt 2$. From \cite{UoLuMoHe16, AnKu20} we know that the optimal value is 
$\lambda \leq \frac{1}{M \sin(\pi/(2M))},$
so the Fisher information criterion is optimal iff $M=2$.

%%%%%%%%%%%%%%
\subsection{Three unbiased dichotomic qubit measurements}\label{subsec:three-qubit-obs}
%%%%%%%%%%%%%%

\begin{figure}[htbp]
    \centering
    \includegraphics[scale=0.35]{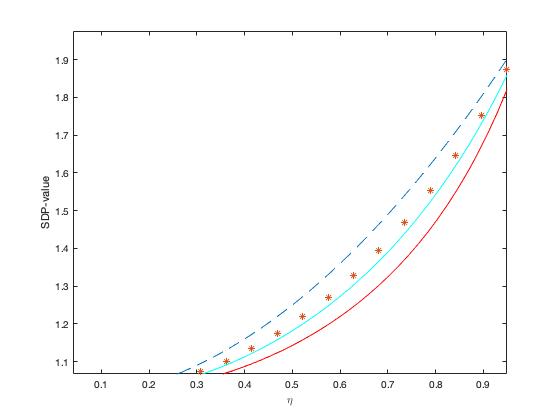}
   \quad
    \includegraphics[scale=0.35]{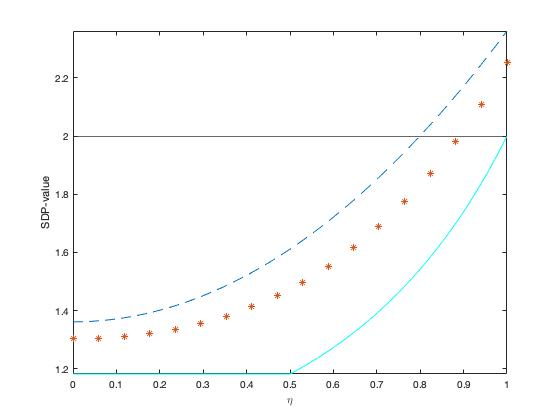}
   \quad 
    \includegraphics[scale=0.35]{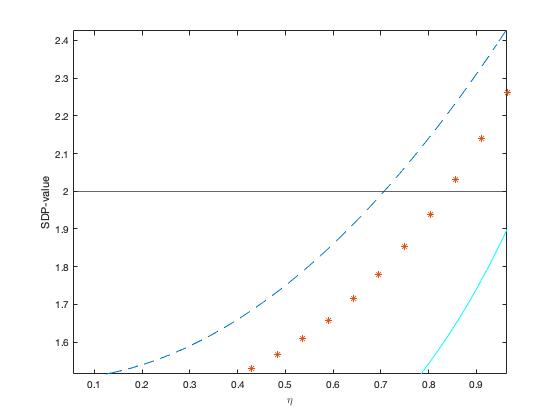}
    \quad 
    \includegraphics[scale=0.35]{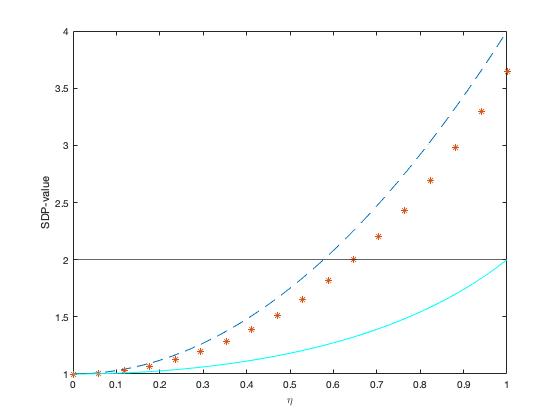}
    \caption{The plot of the height functions $h(v_1,v_2,v_3,\eta,0,0)$ (first), $h(v_1,v_2,v_3,\eta,1/2,1/3)$ (second)  $h(v_1,v_2,v_3,\eta,1/2,1/2)$ (third) and  $h(v_1,v_2,v_3,\eta,\eta, \eta)$ (fourth)  of the
     generalized-Zhu matrix corresponding to three unbiased dichotomic qubit measurements ( $\sigma_x,$ $\sigma_y$, and $\sigma_z$) for  $\rho=\frac{1}{2}(I+v\cdot \sigma)$; $v=(0,0,0)$ is dashed ($--$) line, $v=(1/2,1/3,1/3,0)$ is star ($*$) line, $v=(1/\sqrt{2},0,1/\sqrt{3})$  the read line and $v(1/\sqrt{3},1/\sqrt{3},1/\sqrt{3})$ for cyan line.}
    \label{fig:XYZ}
\end{figure}

\begin{figure}[htbp]
    \centering
    \includegraphics[scale=0.45]{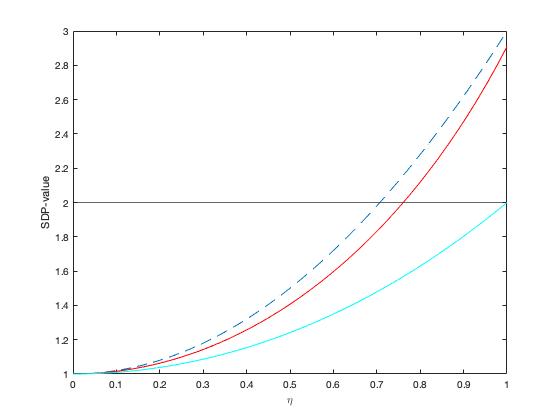}
    \caption{The plot of the height functions $h(v_1,v_2,v_3,\eta,\eta, \eta)$ of the
     generalized Fisher information map corresponding to three unbiased dichotomic qubit measurements corresponding to $\eta_1=(0,1,0)$ and $\vec{n}_k=(\cos \frac{k-1}{3}\pi, \sin \frac{k-1}{3}\pi,0), k=2,3 $, for  $\rho=\frac{1}{2}(\id +v\cdot \sigma)$; $v=(0,0,0)$ is dashed ($--$) line, $v=(1/2,1/3,1/3)$ is red line  }
    \label{fig:X-Planar}
\end{figure}

\begin{figure}[htbp]
    \centering
    \includegraphics[scale=0.45]{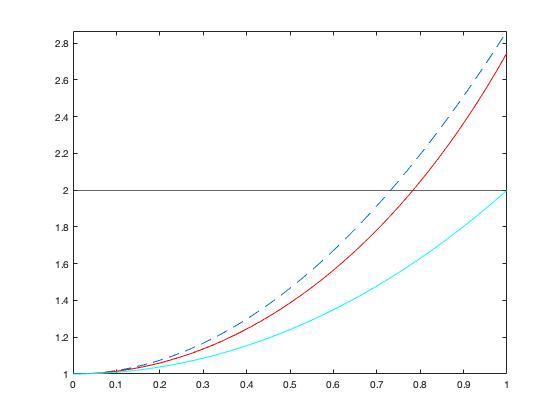}
    \caption{The plot of the height functions $h(v_1,v_2,v_3,\eta,\eta, \eta)$ of the
     generalized Fisher information map corresponding to three unbiased dichotomic qubit measurements corresponding to $\eta_1=(0,0,1)$ and $\vec{n}_k=(\cos \frac{k-1}{3}\pi, \sin \frac{k-1}{3}\pi,0), k=2,3 $, for  $\rho=\frac{1}{2}(I+v\cdot \sigma)$; $v=(0,0,0)$ is dashed ($--$) line, $v=(1/2,1/3,1/3)$ is red  line.}
    \label{fig:ZPlanar}
\end{figure}

We consider now the case of three unbiased dichotomic qubit measurements given by $\A_k(\pm1)=\frac{1}{2}(\id \pm\eta_k\vec{n}_k\cdot \sigma),\,  k=1,2,3$, where $\eta_k\leq 1$ is the    sharpness.
Necessary and sufficient condition for joint measurability of these measurements is presented in \cite{YuOh13} in terms of Fermat-Torricelli (FT) point $\vec{v}_{FT}$ (the point in $\mathbb{R}^3$  that minimizes the sum of the  distances from itself to a given set of points). It holds that the measurements $\A_k$, $k=1,2,3$, are triplewise jointly measurable if and only if 
\begin{equation}\label{cond:FT}
    \sum\limits_{i=0}^3\|\vec{v}_{FT}-\vec{v}_i\|\leq 4
\end{equation}
   where $\vec{v}_{FT}$ is FT point of the following four points in $\mathbb{R}^3$: $\vec{v}_0=-\sum\limits_{k=1}^3\eta_k\vec{n}_k$
 and  $\vec{v}_k=-2\eta_k \vec{n_k}-\vec{v}_0,\, k=1,2,3$. The FT point of four points does not have an analytical expression, except the case of coplanar vectors and  the case when one of the vector is orthogonal to other two vectors.
 \par In the following  we  make a  comparison (see  Table \ref{tb:1}) between the necessary and sufficient compatibility condition given by the FT-point and Fisher information  necessary incompatibility criterion, given by Theorem \ref{thm:Zhu-criterion}. We  compare the  compatibility region derived from the analytical condition of the FT point, with the one obtained by the Fisher information criterion (we use the Matlab \texttt{cvx} package for computing the height of Fisher information map,  $\mathfrak h( \Fim(\A_k)))\, k=1,2,3$, introduced  by eq.~\eqref{eq:def-h}).
 In the first line of the table \eqref{tb:1} we present the case of trine spin measurements, that is the case of three measurements (of global sharpness $\eta$) in an equatorial plane of the Bloch ball equiangularly separated by an angle of $\pi/3$, that is the case $\vec{n}_k=(\cos \frac{k-1}{3}\pi, \sin \frac{k-1}{3}\pi,0), k=1,2,3$. The coplanar property of the vectors allows to derive explicitly the FT point and consequently, the necessary and sufficient condition for compatibility admits an analytical form \cite{YuOh13, AnKu20}
  \begin{equation}
      \eta(\|\vec{n}_1+\vec{n}_2\|+\|\vec{n}_1-\vec{n}_3\|+\|\vec{n}_2-\vec{n}_3\|)\leq 2
  \end{equation}
   We notice that Fisher information condition of incompatibility $\eta>\frac{1}{\sqrt{2}}$ is weaker by confront to the compatibility condition given by FT point, that is $\eta\leq 2/3$. This results confirms again the results presented at example  \eqref{subsec:planar-obs}, where we claim that  Fisher information incompatibility condition is optimal for two planar orthogonal qubit measurements.
   Another interesting case is when one of the vectors, $\eta_1$, is orthogonal on the other two ($\vec{n}_1\perp \vec{n}_2, \vec{n}_3$). More exactly, we set  $\vec{n}_1=(0,0,1),\vec{n}_k=(\cos \frac{k-1}{3}\pi, \sin \frac{k-1}{3}\pi,0), k=2,3 $. In this case, the FT point can be found explicitly and  the corresponding compatibility condition \eqref{cond:FT} becomes:
    \begin{equation}
      \eta(\|\vec{n}_2+\vec{n}_3\|+\|\vec{n}_2-\vec{n}_3\|)\leq 2\sqrt{1-\eta^2\|\vec{n}_1\|^2},
  \end{equation}
  and setting the compatibility region for $\eta\leq 0.59$.
  On the other hand  Zhu  incompatibility condition requires $\eta>0.72$.
    
   \begin{table}[htpp]
\begin{center}
\bgroup
\def\arraystretch{1.5}
\begin{tabular}{ |c| c| c| c|}\hline
 spin measurements & Fisher inf. incomp. cond.& FT point  comp. cond.  \\  \hline\hline
coplanar, equiangular vectors & $\eta>\frac{1}{\sqrt{2}}\approx 0.7071$ & $\eta<\frac{2}{3}\approx 0.6667$  \\  \hline
  one vector orth.~to the other two vectors & $\eta>0.72$ & $\eta<0.5907$   \\
 \hline
\end{tabular}
\egroup
\end{center}
\caption{Comparison  between Fisher information incompatibility condition and FT necessary and sufficient compatibility condition}
\label{tb:1}
\end{table}
A good perspective about the incompatibility region
given by the Fisher information criterion for
three unbiased dichotomic qubit measurements can be found using  using the generalized Fisher map.
It is known that, if the vectors $\vec{n}_k,k=1,2,3$ are mutually orthogonal directions, then the condition \eqref{cond:FT} reduces to the sufficient and sufficient  condition for joint measurability, that is 
$$\eta_1^2+\eta_2^2+\eta_3^2\leq 1$$
In the following graphic we aim to analyze the influence of the state $\rho$ on determining the incompatibility region for three unbiased dichotomic qubit measurements, that correspond to $\sigma_x,\sigma_y$ and $\sigma_z$.
This case is presented in Figure \ref{fig:XYZ} using the computation of the height function
$\mathfrak h\left( \Fimrho(\mathcal{A}) \right)\equiv
h(v_1,v_2,v_3,\eta_1,\eta_2,\eta_3)$ which is now a function of the sharpness $\eta_i, i=1,2,3$ and of the parameters of the Bloch vector $v_i,i=1,2,3$.

The simplest case for $\eta_1=\eta_3=0$ that correspond to compatible measurements, for any $\eta_2\in[0,1]$, is confirmed by the graphic from Figure \ref{fig:XYZ}, as the curves are below the line $y=2$, which delimits the compatibility and the incompatibility regime.
In the setting $\eta_2=1/2$ and $\eta_3=1/3$, we know that the compatibility condition is  $\eta_1\leq 0.79$, where as for $\eta_2=\eta_3=1/2$, we know that the compatibility condition is  $\eta_1\leq 0.7$. 
We plot in Figure \ref{fig:XYZ} these two cases (the second and the third graphic), and for different values of the Bloch vector $v$.
As can be seen, in both cases,  the incompatibility regime (above $y=2$) is recovered "faster" for the case of the Bloch vector $v=(0,0,0)$, which corresponds to the classical Fisher information map. 
Moreover, the plots confirm  that we cannot detect incompatibility regimes using unit Bloch vector. Indeed, the cyan line that corresponds to the height of the Fisher matrix is always below 2, in this case.

In Figures \ref{fig:X-Planar} and \ref{fig:ZPlanar} we present that the cases  of coplanar POVMs, for which we have compared the Zhu incompatibility region and the compatibility region found by FT point. We see that in both cases the best way to detect incompatibility is to use the maximally mixed state, whereas $\rho$ being pure cannot detect incompatibility.

We leave open the existence of a set of measurements which can be shown to be incompatible using the generalized Fisher information map $\Fim_\rho$ for $\rho \neq \id$, but for which the standard map $\Fim$ cannot detect incompatibility. The set of examples we considered in this section show that in most cases of interest, the standard map $\Fim$ performs better that its shifted variant $\Fim_\rho$.

%%%%%%%%%%%%%%%%%%%%
\section{Conclusions and perspectives}
%%%%%%%%%%%%%%%%%%%%

The post-processing order of quantum measurements is a natural and important relation, describing intrinsic noise in measurements and giving an operational definition for quantum incompatibility. We have presented a general framework where the set of (equivalence classes of) measurements is mapped into matrices and the post-processing order translates to the usual matrix ordering. We have illustrated that it depends on the chosen order morphism how much information is lost. Some reasonably looking maps loose too much information and cannot give any criteria for incompatibility, even if they can still detect some non-ordering. In the ideal case, one would get rid of uninteresting details while keeping enough essential information to, e.g., decide incompatibility with high accuracy. The Fisher information map introduced by Zhu is an example of a good and useful order preserving map. It is indeed optimal among quadratic maps as we have shown.

The presented general setting indicates directions for future studies. Firstly, one can seek modifications of the Fisher information map and try to find efficient incompatibility tests by using them. For example, the generalized Fisher information map is such a generalization which could be used to detect incompatibility where the usual map cannot. Instead of using the identity matrix, one could use other quantum states $\rho$. Full rank states should lead to interesting criteria, since the incertitude about a lower rank state is smaller, and thus the Fisher information matrix should be less informative. Moreover, we have shown that unit rank states are useless for incompatibility detection, so invertible states $\rho$ should be investigated. On the other hand, most quantum measurements of interest have some kind symmetry; the most informative state $\rho$ should also enjoy the same symmetry, making the choice $\rho = \id / d$ reasonable. We leave the question of the hierarchy between the different Fisher information maps $\Fim_\rho$ open for further investigation. It is also possible to investigate order preserving maps that are not quadratic. 

Secondly, the underlying idea of order preserving maps that simplify the structure while keeping essential features can be used for other quantum devices than just for quantum measurements. In particular, preparators and instruments have their corresponding post-processing orders and one can formulate analogous maps for them.
There, again, the aim would be to forget uninteresting details but keep the essential structure leading to physically relevant phenomena such as incompatibility.

\bigskip

\emph{Acknowledgements.} I.N.~ was supported by the ANR project \href{https://esquisses.math.cnrs.fr/}{ESQuisses}, grant number ANR-20-CE47-0014-01.

\newcommand{\etalchar}[1]{$^{#1}$}

\end{document}